\documentstyle[12pt,titlepage]{article}

\font\grande=cmr10 scaled \magstep4
\font\medio=cmr10 scaled \magstep2
\outer\def\beginsection#1\par{\medbreak\bigskip
      \message{#1}\leftline{\bf#1}\nobreak\medskip\vskip-\parskip
      \noindent}

\def\laq{\raise 0.4ex\hbox{$<$}\kern -0.8em\lower 0.62
ex\hbox{$\sim$}}
\def\gaq{\raise 0.4ex\hbox{$>$}\kern -0.7em\lower 0.62
ex\hbox{$\sim$}}

\def \ra {\rightarrow}

\def \H {{a^\prime \over a}}

\def \da {\delta}
\def \la {\lambda}

\def \Da {\Delta}
\def \b {\beta}
\def \a {\alpha}

\def \noi {\noindent}

\def\sqr#1#2{{\vcenter{\hrule height.#2pt\hbox{\vrule width.#2pt
height#1pt \kern#1pt\vrule width.#2pt}\hrule height.#2pt}}}

\def\lsim{\mathrel{\rlap{\lower4pt\hbox{\hskip1pt$\sim$}}
    \raise1pt\hbox{$<$}}}         %less than or approx. symbol
\def\gsim{\mathrel{\rlap{\lower4pt\hbox{\hskip1pt$\sim$}}
    \raise1pt\hbox{$>$}}}         %greater than or approx. symbol

\def\bl{\Biggl\{}
\def\br{\Biggr\}}

\def\O{{\cal O}}
\def\a{\alpha}
\def\b{\beta}
\def\d{\delta}

\def\D{{\Delta}}

\def\F{{\cal F}}

\def\H{{\cal H}}
\def\g{\gamma}

\def\n{\eta}

\def\s{\sigma}
\def\T{\Theta}

\def\vphi{\varphi}

\def\hf{\frac{1}{2}}
\def\der{\partial}
\def\bq{\begin{equation}}
\def\eq{\end{equation}}
\def\brr{\begin{eqnarray}}
\def\err{\end{eqnarray}}
\def\ba{\left(\begin{array}}
\def\ea{\end{array}\right)}

\def\ba{\left(\begin{array}}
\def\ea{\end{array}\right)}
\begin{document}
\bibliographystyle {unsrt}
\newcommand{\pa}{\partial}
\newcommand{\rhob}{{\bar \rho}}
\newcommand{\prb}{{\bar p}}

\titlepage
\begin{flushright}
CERN-TH.7544/94 \\
\end{flushright}
\vspace{5mm}
\begin{center}
{\grande Metric Perturbations in Dilaton-Driven Inflation}

\vspace{5mm}

R. Brustein, M. Gasperini\footnote{Permanent address: {\em
Dipartimento di Fisica Teorica,
Via P. Giuria 1, 10125 Turin, Italy.}}, M. Giovannini \\
{\em Theory Division, CERN, CH-1211 Geneva 23, Switzerland} \\
V. F. Mukhanov \\
{\em Institut fur Theoretische Physik ETH,
CH-8093 Zurich, Switzerland} \\
and \\
 G. Veneziano \\
{\em Theory Division, CERN, CH-1211 Geneva 23, Switzerland} \\
\vspace{5mm}
{\medio  Abstract} \\
\end{center}
\noi
We compute the spectrum of scalar and tensor metric
perturbations generated, as amplified vacuum fluctuations,
during an epoch of dilaton-driven inflation of the type
 occurring naturally in string cosmology. In the tensor case
the computation is straightforward while,
 in the scalar case, it is made delicate by
 the appearance of a growing mode
in the familiar longitudinal gauge. In spite of this, a reliable
perturbative calculation of perturbations far outside
the horizon can be performed by resorting either to
appropriate  gauge invariant variables, or to
a new  coordinate system in which the growing mode
can be ``gauged down". The simple outcome of this complicated
analysis is that both scalar and tensor perturbations exhibit nearly
Planckian spectra, whose common ``temperature" is related to some very
basic
parameters of the string-cosmology background.
\vspace{8mm}
\vfill
\begin{flushleft}
CERN-TH.7544/94 \\
December 1994
\end{flushleft}

\newpage
\

\renewcommand{\theequation}{1.\arabic{equation}}
\setcounter{equation}{0}
\section {Introduction}

According to quantum mechanics,
   metric and  energy-density fluctuations
are necessarily present as tiny wrinkles on top of any, otherwise
homogeneous, classical cosmological background.
 It is well known \cite{gris} that
transitions from one cosmological era to another may lead to a
parametric amplification of such perturbations, which eventually
reveal themselves as stochastic classical inhomogeneities.
In particular, ``slow rolling" scenarios leading
to de-Sitter like inflation \cite{inflation}
predict an almost scale-invariant spectrum of density (scalar)
perturbations \cite{first,second} and of gravitational waves
\cite{star} (for a review, see ref.\cite{MFB}).

In string cosmology, inflation is expected to
be associated with a phase of growing curvature \cite{GSV}
and dilaton coupling  \cite{Ven1} (called
``pre-big-bang" scenario in \cite{GV1}), in which the
accelerated evolution
of the scale factor, $a(t)$, is driven by the kinetic energy
of the
dilaton field,
with negligible contributions from the dilaton potential
\cite{GV1}-\cite{RAMY} (see also \cite{behrndt}, and \cite{Levin} for
related, though differently motivated,
issues in the context of scalar-tensor cosmologies).
This inflationary phase is most naturally described in the string
frame (S-frame, also called, somewhat improperly,  Brans-Dicke
frame),
 in which weakly coupled strings move along geodesic surfaces
\cite{SV}.
 In the S-frame,
isotropic solutions of the string cosmology equations
describe an accelerated expansion of the ``pole-inflation" type
\cite{sha}, i.e. one characterized by $\dot a >0$, $\ddot a >0$,
$\dot H >0$,
where $H=\dot a/a$, and a dot denotes differentiation with respect to
cosmic time $t$. In the conformally related Einstein frame
(E-frame), in which the
curvature term
and dilaton kinetic term in the action are diagonalized in
the standard canonical form, the corresponding
isotropic solutions describe instead an accelerated contraction
\cite{GV2,GV3,GAS}, characterized by
$\dot a <0$, $\ddot a <0$, $\dot H <0$. In the E-frame the scale
factor
can be parametrized, in conformal time $\eta$ ($dt \equiv a d \eta$),
as
\bq
a \sim (-\eta)^{\alpha},~~~~~~~ \alpha>0, ~~~~~~
\eta\rightarrow 0_{-}.
\label{1}
\eq
(see \cite{GV2,GV3,GAS} for a discussion of how the standard
kinematical
problems are solved in such a contracting background).
We recall, for future reference, that the solution with $\a =\hf$
corresponds to a pure four-dimensional dilaton-dominated background,
while the case $\a >\hf$ occurs, in four dimensions, in the
presence of additional string matter sources \cite{GV2,GV3}.

The epoch of accelerated evolution is assumed to end
 \cite{GSV},\cite{Ven1},\cite{GV1} at some time $|\eta| =
\eta_1$ when
a maximal curvature scale $H_1\equiv H(\eta_1)$ is reached, i.e. when
higher-derivative terms
in the  string effective action become important. Both in the S-frame
and in the E-frame that
point is
reached  when
$a_1\eta_{1} =\O(\lambda_{s})$, where $\lambda_{s}\simeq
\sqrt{\a' \hbar}$ is the fundamental
length of string theory. In the S-frame $\lambda_{s}$
is a constant and the
Planck length $\lambda_{p}$ is given by $\lambda_p =
 g_{string} \lambda_{s} = e^{\hf \varphi} \lambda_s$
($\varphi$ is the dilaton field), while in
the E-frame the fundamental constant is $\lambda_{p}$ (and
$\lambda_s =
 e^{-\hf \varphi} \lambda_p$). In any case,
 $H_1 \laq M_p$ ($M_p=G^{-1/2}$)  if
inflation ends when the dilaton is still in the perturbative regime.
 A smooth transition to standard cosmology at
the end of the accelerated pre-big-bang
evolution is expected to be controlled both by
$\a'$ corrections to the low energy effective action and by the
contribution of a non-perturbative dilaton potential, as discussed in
\cite{GV4} (recent related work concerning the possible smoothing out
of
curvature singularities in string theory
can be found
in \cite{KK}, \cite{ts} and  \cite{martinec}).

The accelerated
shrinking of the event horizon during inflation \cite
{GV1}-\cite{GV3}, \cite{GAS}, and the subsequent transition to
a decelerated, radiation-dominated background, produces a dramatic
amplification of the initial vacuum fluctuations since,
in an inflationary background of the
pre-big-bang type, the comoving amplitude of metric
perturbations outside the horizon, instead
of remaining constant, grows asymptotically \cite{GV2}.

This peculiar effect can be easily illustrated
by considering the evolution of tensor metric
perturbations in
the E-frame, where the background scale factor is given by
eq. (\ref{1}), and
each Fourier mode of the perturbation satisfies,  to
lowest
order,
the simple equation \cite{gris}
%%%%%%%%%%%%%%%%%%%%%%%%%%%%%%%%%%%%%%%%%%%%%%%%%%%%%%%%%%
\bq
h''_k+2 \frac{a'}{a}h'_k+k^2 h_k=0 \; ,
\label{grav}
\eq
%%%%%%%%%%%%%%%%%%%%%%%%%%%%%%%%%%%%%%%%%%%%%%%%%%%%%%%%%%
(a prime denotes differentiation with respect to $\eta$).
The asymptotic solution of eq. (\ref{grav}) well
outside the horizon ($|k\eta|<<1$) is given
by
%%%%%%%%%%%%%%%%%%%%%%%%%%%%%%%%%%%%%%%%%%%%%%%%%%%%%%%%%%
\bq
h_k=A_k +B_k \int^\eta{d\eta' \over a^2(\eta')}=A_k +B_k
\int^\eta d\eta'(-\eta')^{-2\a}  \; ,
\label{GRAVSOL}
\eq
%%%%%%%%%%%%%%%%%%%%%%%%%%%%%%%%%%%%%%%%%%%%%%%%%%%%%%%%%%
where $A_{k}$, $B_{k}$ are integration constants.
For $\a<\hf$,  $h_k$ approaches a constant asymptotically.
 The typical amplitude  of  fluctuations over scales $k^{-1}$,
normalized to
an initial vacuum-fluctuation spectrum, is given as usual
 by \cite{star,MFB}
\bq
|\delta_{h_k}|=k^{3/2}|h_k| \simeq \left(H\over M_p\right)_{HC}
\simeq {H_1\over M_p} (k\eta_1)^{1+\a}
\eq
(the subscript ``HC" on a time
dependent quantity means that it is to be evaluated at the time of
horizon
crossing for the particular scale $k^{-1}$ under consideration, i.e.
at
$|\eta|=\eta_{HC}\simeq k^{-1}$).

If, on the contrary, $\a \geq \hf$
(which is indeed the case for ``realistic" solutions of the string
cosmology equations \cite{GV1,GV2,GV3}), then the asymptotic solution
(\ref{GRAVSOL}) is dominated by the growing mode, and the typical
amplitude of tensor perturbations over scales $k^{-1}$ varies in time
according to
\bq
|\delta_{h_k}(\eta)|=k^{3/2}|h_k| \simeq {k \eta \over a^2 M_p}
\left(a \over \eta \right)_{HC} \simeq
\left(H\over M_p\right)_{HC}\left(a_{HC}\over a \right)^2 k \eta
\simeq {H_1\over M_p} (k\eta_1)^{1+\a} |k\eta|^{1-2\a}
\label{spectrum}
\eq
with an additional $\ln |k\eta|$ factor appearing at $\a= \hf$.
As we shall discuss
in Sec.2 (see also \cite{GV2,GV3}), the same result is obtained
in the conformally related S-frame  in
which the background
metric describes accelerated expansion instead of contraction,
and eq. (\ref{grav}) is
modified
by an explicit coupling of the perturbation to the
time variation of the
dilaton
background \cite{GG1}. We stress that
the growth of the comoving amplitude of tensor perturbations can
be
understood as a consequence of the joint contribution of the metric
and
of the
dilaton background to the ``pump" field responsible for the
parametric
amplification process
$\cite{gris1}$, and is thus  to be expected, in general, in case
of perturbations evolving in scalar-tensor backgrounds, as
noted also in \cite{Barrow}.

The final amplitude $|\da_{h_k}(\eta_1)|$ thus depends on the
power $\a$ which characterizes the background. In this paper we shall
 concentrate on the case $\a=\hf$ which corresponds
to a purely dilaton-driven isotropic inflation in $3+1$ dimensions.
 From the point of view of
string theory, neglecting everything but the dilaton
 is particularly appealing, since it
corresponds to taking a Conformal-Field-Theory as the starting
homogeneous
background.
Furthermore,  even if a diluted gas of classical
strings is added in the initial conditions,   its effect is simply to
ignite an accelerated evolution of the flat perturbative vacuum
towards the dilaton-driven inflationary
regime
\cite{GV3,GAS}. The  matter contribution becomes eventually
negligible and the scale factor ends up evolving  (in the E-frame)
as $|\eta|^{1/2}$.

The case $\a=\hf$
does not pose any problem for tensor perturbations since, according
to
eq. (\ref{spectrum}), the condition $|\da_{h}(\eta_1)|\laq 1$ is
 satisfied for all $\a \leq 2$ (provided $H_1\laq M_p$), at all
scales $k^{-1}>\eta_1$ (smaller scales are not parametrically
amplified).
The situation appears to be drastically different for
scalar
perturbations, which become instead too large asymptotically
to be consistent
with the
usual description in terms of the linearized gauge-invariant
formalism
\cite{Bardeen,MFB}.
Consider indeed
the canonically normalized field $v_k$
associated
with scalar perturbations (see Sect. 3).
The variable $v/a$ obeys again eq.(\ref{grav}), hence behaves
asymptotically
as in (\ref{GRAVSOL}).
Given the relation between $v$
and the
scalar metric perturbation
$\psi$ in the longitudinal gauge, one finds
for the typical amplitude of $\psi$  over scales
$k^{-1}>>\eta$ and  at
 time $\eta$ (Section 3):
%%%%%%%%%%%%%%%%%%%%%%%%%%%%%%%%%%%%%%%%%%%%%%%%%%%%%%%%%%
\bq
    \left|\d_{ \psi_k}(\n)\right| \simeq
\frac{k}{M_p |k\n|^2} \frac{\n}{a^2}
\left(\frac{a}{\n}\right)_{\rm HC}
\simeq \frac{H}{M_p}\frac{a_{\rm HC}}{a}\simeq\frac{H}{M_p}
|k\n|^{-1/2}
\eq
%%%%%%%%%%%%%%%%%%%%%%%%%%%%%%%%%%%%%%%%%%%%%%%%%%%%%%%%%%
 For any given $\eta$, there is thus a
frequency band defined by the condition
 $k< \eta^{-1} [H/M_{P}]^{2}$, for which  $ | \delta_\psi |>1$,
and  the
perturbative approach apparently breaks down.
Alternatively, at sufficiently small $k$,
the (naive) spectral energy density evaluated at the
end of inflation (i.e. at the beginning of radiation-dominance)
is larger than  critical,
%%%%%%%%%%%%%%%%%%%%%%%%%%%%%%%%%%%%%%%%%%%%%%%%%%%%%%%%%%
\bq
\Omega_k(\n_1)=\frac{k}{\rho_c(\n_1)}\frac{d\rho}{d k}\simeq
\left|\d _{\psi_k}(\n_1)\right|^2\simeq
\left(\frac{H_1}{M_p}\right)^2
\left(\frac{k_1}{k}\right)>1 ,
\eq
%%%%%%%%%%%%%%%%%%%%%%%%%%%%%%%%%%%%%%%%%%%%%%%%%%%%%%%%%%
($k_{1}\equiv 1/\eta_{1}$) in contrast with the  hypothesis
of a negligible back-reaction of the perturbations on the background
 metric (incidentally, a similar problem was found to arise for
scalar
perturbations in the context of Kaluza-Klein cosmologies, as a
consequence
of the shrinking internal dimensions \cite{abbott}, but was left, to
the best of
our knowledge, unsolved).

Unless the  inflationary growth of the background stops at a
sufficiently small
 Hubble parameter $H_1$, scalar
perturbations apparently do not remain small.
In a
string theory context, however,
the curvature scale at the end of inflation is expected to be the
string scale \cite{GV1}-\cite{GV4}, $H_1\sim \la_s^{-1}$.
Consequently, a very small value of the string
coupling  would be required at the end of inflation. Otherwise, a
full
non-linear approach would  seem to be required in order to follow the
evolution
of scalar perturbations in such a dilaton-dominated
background, and to make
predictions about their final spectrum.

The main purpose of this paper is to show that new appropriate
perturbative
techniques can be developed in order to follow the evolution
of such ``large" perturbations throughout the inflationary epoch.
This will allow us to argue that, because of its special properties,
the growing mode does not lead to an inhomogeneous Universe
 at least if one starts with minimal (i.e. vacuum quantum)
primordial fluctuations.
In support of this claim we will present explicit first and second
order
calculations performed in a different -- and so far to our knowledge
 unexploited -- ``off-diagonal"
gauge. The  results clearly show that perturbation theory
does not break down in the new gauge, that inhomogeneities remain
bounded, and that their spectrum can be computed reliably.
We shall arrive at similar conclusions by using  a different approach
based on
appropriate covariant and gauge-invariant variables \cite{EB,BE}.
The spectral amplitude of density fluctuations turns out to coincide,
 quite unexpectedly,  with
previous results obtained  in the longitudinal gauge \cite{GV3}
by neglecting (with a dubious argument) the growing mode
contribution to the scalar perturbation amplitude.

The paper is organized as follows. In Section 2 we recall,
for completeness and later comparison,
previous works on tensor perturbations in a string cosmology context,
and extend  it to
the case of a dilaton-driven background with
extra compactified dimensions. We
 stress the emergence of tilted spectra,
favoring shorter scales, and  the stability of the spectrum
 with respect to the choice of the background solution.
In Section 3 we show how the presence of a growing solution for the
scalar
components of the metric and dilaton perturbations  in the
longitudinal
gauge
invalidates a perturbative analysis, typically at small
wave numbers and towards the end of the inflationary epoch.
 We show, in the Appendix, that
the growing asymptotic solution can  neither be eliminated by an
appropriate choice of the number of spatial dimensions, nor by
considering anisotropic, Bianchi I type metric backgrounds with
an arbitrary number of shrinking internal dimensions.
In Section 4, we  abandon momentarily
the metric perturbation approach in favour of the fluid flow approach
pioneered
by Hawking \cite{HAW}, extensively applied by Liddle and Lyth
\cite{LL}
and
more recently developed by Bruni and Ellis \cite{BE}.
We show that such variables allow for a consistent direct
computation of the spectral energy density of the metric and
dilaton fluctuations, without
 any sign of breakdown of the linear approximation.
Computing the size of second
order corrections directly
in these variables looks, however, too difficult a task.
Armed with the knowledge that physical observables, such as the
energy
density stored in the perturbations, remain small, we look, in
Section 5,
for a more suitable gauge choice incorporating this feature.
 And, indeed, we are able to identify
 a new reference frame in which   the
growing mode at $k=0$ is ``gauged away",  the small-$k$ growing modes
are ``gauged down", and a reliable perturbative scheme can be
developed.
All this is confirmed
by  an explicit calculation of the relevant second
order quadratic terms, which allow us to give an estimate of the size
of the second order corrections to metric perturbations.
 Our main conclusions  are finally summarized in Section 6.

\renewcommand{\theequation}{2.\arabic{equation}}
\setcounter{equation}{0}
\section{Tensor perturbations}

In this Section we recall the main characteristics
of tensor perturbations in a dilaton-dominated
background, stressing in particular
their
stability against the addition of extra dimensions or the choice of
different,
duality-related solutions of the string cosmology equations.

In the  E-frame, the equations obtained from the low energy string
effective action \cite{lov}, for a torsionless background, are simply
given by
%%%%%%%%%%%%%%%%%%%%%%%%%%%%%%%%%%%%%%%%%%%%%%%%%%%%%%%%%%
\bq
R_{\mu}^{\ \nu}-\hf \d_{\mu}^{\ \nu} R=
\hf\left( \der_\mu\vphi\der^\nu\vphi-
\hf\d_{\mu}^{\ \nu} \der_\a\vphi\der^\a\vphi\right)
\label{einstein}
\eq
%%%%%%%%%%%%%%%%%%%%%%%%%%%%%%%%%%%%%%%%%%%%%%%%%%%%%%%%%%
%%%%%%%%%%%%%%%%%%%%%%%%%%%%%%%%%%%%%%%%%%%%%%%%%%%%%%%%%%
\bq
 g^{\a\b}\nabla_\a \nabla_\b\vphi=0
\label{dilaton}
\eq
%%%%%%%%%%%%%%%%%%%%%%%%%%%%%%%%%%%%%%%%%%%%%%%%%%%%%%%%%%
 where $\vphi$ is the dilaton field and, unless
otherwise stated, we shall adopt units in which
$16\pi G=1$. It is well known that eq.(\ref{dilaton}) is a
consequence
of eq.(\ref{einstein}), to which we shall therefore restrict our
attention from now on.
Note that  we  neglect the contribution
of a possible dilaton self-interaction potential having in mind that
the
whole evolution starts out in the weak coupling region.
Looking for spatially flat solutions in which there are $d$
spatial
dimensions which evolve in time with a scale factor $a(\n)$, while
other $n$ internal dimensions simultaneously
shrink with a scale factor $b(\n)$,
$$
g_{\mu\nu}= diag(a^2, -a^2 \d_{ij}, -b^2\d_{mn}),~~~~~~~
\varphi =\varphi (\eta)
$$
\bq
i,j=1,\cdots,d ~~~~~~~~~~~~~~~~~~~~ m,n=d+1,\cdots,d+n
\label{ani}
\eq
the Einstein equations (\ref{einstein}), (\ref{dilaton}) take the
explicit form
\brr
d(d-1)\H^2+n(n-1) {\cal F}^2+2n d \H {\cal F}&=&\hf \vphi'^2\nonumber
\\
2(d-1) \H'+(d-1)(d-2)\H^2+2n {\cal F}'+n(n+1) {\cal F}^2+ 2n (d-2) \H
{\cal F}
&=&-\hf \vphi'^2 \nonumber \\
2(n-1){\cal F}'+2d\H'+d(d-1)\H^2+n(n-1) {\cal F}^2+ 2(d-1)(n-1) \H
{\cal F}
&=&-\hf \vphi'^2 \nonumber \\
\vphi''+[(d-1)\H +n {\cal F}]\vphi'&=&0
\err
where $ \H=a'/a = aH$ and ${\cal F}= b^{\prime}/b$. We shall
consider,
in particular, the exact anisotropic solution parametrized, for $\eta
\ra 0_-$, by
$$
a=(-\n)^\a,~~~~~~~~~~b=(-\n)^\b,~~~~~~~~~~
\vphi=\frac{n-d-\sqrt{d+n}}{1+\sqrt{d+n}}\ln(-\n)+const
$$
\bq
\a=\frac{\sqrt{d+n}+1-2n}{(1+\sqrt{d+n})(d+n-1)}, ~~~~~~~~~~~~~~~~
\b=\frac{\sqrt{d+n}-1+2d}{(1+\sqrt{d+n})(d+n-1)}
\label{ddimback}
\eq
Such a background is a particularly significant
candidate for
describing a phase of inflation plus
dynamical dimensional reduction in a string cosmology context.
Indeed, if one goes over
to the S-frame  by the conformal transformation:
%%%%%%%%%%%%%%%%%%%%%%%%%%%%%%%%%%%%%%%%%%%%%%%%%%%%%%%%%%
\bq
\widetilde g_{\mu\nu}^{string}=g_{\mu\nu}\
e^{\hbox{$\frac{2\vphi}{d+n-1}$}},
\label{confor}
\eq
%%%%%%%%%%%%%%%%%%%%%%%%%%%%%%%%%%%%%%%%%%%%%%%%%%%%%%%%%%
one finds a particular case of the general exact dilaton-driven
solution
in critical dimensions \cite{Ven1,GV3,Muller,MV},
in which ``external" and ``internal"  scale factors
are
related by the duality transformation $\tilde b = 1/{\tilde a}$.

Each Fourier mode $h_{k}$ of the transverse--traceless tensor
perturbations
 $\delta g_{ij}=- a^{2} h_{ij}(\eta,\vec x)$
of the ``external" $d$-dimensional metric background
satisfies, in the E-frame, the free scalar field equation \cite
{gris,GG1,dem}
%%%%%%%%%%%%%%%%%%%%%%%%%%%%%%%%%%%%%%%%%%%%%%%%%%%%%%%%%%
\bq
h''_k+\left[(d-1)\H+n\F\right] h'_k+k^2 h_k=0
\eq
%%%%%%%%%%%%%%%%%%%%%%%%%%%%%%%%%%%%%%%%%%%%%%%%%%%%%%%%%%
For the solution (\ref{ddimback}), the coefficient of the
$h_k^{\prime}$
term of
this equation is exactly dimensionality-independent, since
%%%%%%%%%%%%%%%%%%%%%%%%%%%%%%%%%%%%%%%%%%%%%%%%%%%%%%%%%%
\bq
(d-1) \frac{a'}{a} +n\frac{b'}{b}=[(d-1)\a+n\b]\ \frac{1}{\n}
= \frac{1}{\n}
\eq
%%%%%%%%%%%%%%%%%%%%%%%%%%%%%%%%%%%%%%%%%%%%%%%%%%%%%%%%%%
It follows that, in the long wavelength ($|k\eta|\rightarrow 0$)
limit,
%%%%%%%%%%%%%%%%%%%%%%%%%%%%%%%%%%%%%%%%%%%%%%%%%%%%%%%%%%
\bq
h_k=A_k+B_k\ln|k\n| \; .
\label{gravsol}
\eq
%%%%%%%%%%%%%%%%%%%%%%%%%%%%%%%%%%%%%%%%%%%%%%%%%%%%%%%%%%

Tensor perturbations are thus growing logarithmically in a
dilaton-driven inflationary background, quite irrespectively of the
isotropy and
of the number of spatial dimensions. This mild growth, however,
does not prevent a
linearized metric perturbation description of the vacuum
fluctuations,
in any number of dimensions. Consider, in fact, the correctly
normalized
variable $u_{k}$ satisfying canonical commutation relations,
which for
tensor
perturbations in the background (\ref{ani}) is related to $h_{k}$
by
%%%%%%%%%%%%%%%%%%%%%%%%%%%%%%%%%%%%%%%%%%%%%%%%%%%%%%%%%%
\bq
u_k=y h_k,\hspace{.3in} y= a^{(d-1)/2}b^{n/2} \; .
\eq
%%%%%%%%%%%%%%%%%%%%%%%%%%%%%%%%%%%%%%%%%%%%%%%%%%%%%%%%%%
$u_k$ satisfies the equation \cite{GG1,GV2}
%%%%%%%%%%%%%%%%%%%%%%%%%%%%%%%%%%%%%%%%%%%%%%%%%%%%%%%%%%
\bq
u''_k+\left(k^2-\frac{y''}{y} \right) u_k=0
\label{ueq}
\eq
%%%%%%%%%%%%%%%%%%%%%%%%%%%%%%%%%%%%%%%%%%%%%%%%%%%%%%%%%%
with asymptotic solution, for $|k\eta|<<1$,
%%%%%%%%%%%%%%%%%%%%%%%%%%%%%%%%%%%%%%%%%%%%%%%%%%%%%%%%%%
\bq
u_k=c_1 y + c_2 y \int^\n \frac{d\n'}{y^2(\n')}
\eq
%%%%%%%%%%%%%%%%%%%%%%%%%%%%%%%%%%%%%%%%%%%%%%%%%%%%%%%%%%
($c_1$ and $c_2$ are integration constants). Inside the horizon
($|k\eta|>>1$), the amplitude of a freely oscillating, positive
frequency mode, normalized to the initial vacuum state at $\eta=
-\infty$, is represented by $|u_k|\simeq k^{-1/2}$.
Since $y^{2}\sim|\eta|$, we thus obtain for the normalized vacuum
fluctuations
outside of the horizon
%%%%%%%%%%%%%%%%%%%%%%%%%%%%%%%%%%%%%%%%%%%%%%%%%%%%%%%%%%
\bq
 h_k=\frac{u_k}{y}
\simeq
\frac{\ln|k\n|}{\sqrt{k}\ y_{\rm HC}}\; ,
\eq
%%%%%%%%%%%%%%%%%%%%%%%%%%%%%%%%%%%%%%%%%%%%%%%%%%%%%%%%%%
which gives a typical amplitude over scales $k^{-1}$
%%%%%%%%%%%%%%%%%%%%%%%%%%%%%%%%%%%%%%%%%%%%%%%%%%%%%%%%%%
\bq
  \left|\d _{h_k}(\n)\right| \simeq
\left(\frac{H_1}{M_p}\right)^{(d+n-1)/2} (k\n_1)^{(d+n)/2}\ln|k\n|
\label{amplitude}
\eq
%%%%%%%%%%%%%%%%%%%%%%%%%%%%%%%%%%%%%%%%%%%%%%%%%%%%%%%%%%
(we have assumed a final inflation scale $H_{1}$ of the same order as
the final
compactification scale, $H_{1}\simeq (a_{1}\eta_{1})^{-1}\simeq
(b_{1}\eta_{1})^{-1})$. The necessary condition for the
validity of the linear approximation,
$  |\delta_{h}|< 1$,  is therefore satisfied for any $d$, and
for all
scales
$k<k_{1}=1/\eta_{1}$, provided that $H_{1}\laq M_{P}$, i.e. that the
dilaton
$\vphi$ is still in the perturbative region
($e^{\vphi}\laq 1$), at the end
of
inflation. For future reference,  we explicitly write the result for
the $n=0,\
d=3$ case, corresponding to an isotropic four-dimensional  Universe,
%%%%%%%%%%%%%%%%%%%%%%%%%%%%%%%%%%%%%%%%%%%%%%%%%%%%%%%%%%
\bq
 \left|\d _{h_k}(\n)\right| \simeq
\left(\frac{H_1}{M_p}\right) (k\n_1)^{3/2}\ln|k\n| \; .
\label{4damplitude}
\eq
%%%%%%%%%%%%%%%%%%%%%%%%%%%%%%%%%%%%%%%%%%%%%%%%%%%%%%%%%%

We note, finally, that the same results are obtained if tensor
perturbations are linearized in the S-frame, related to the E-frame
by
the
conformal transformation (\ref{confor}). Indeed, in the S-frame, the
dilaton
contribution appears explicitly in the tensor perturbation equation,
which
becomes \cite{GG1}
%%%%%%%%%%%%%%%%%%%%%%%%%%%%%%%%%%%%%%%%%%%%%%%%%%%%%%%%%%
\bq
 h''_k+\left[(d-1)\widetilde\H+n\widetilde {\cal F}-\vphi'\right]
h'_k
+k^2 h_k=0
\eq
%%%%%%%%%%%%%%%%%%%%%%%%%%%%%%%%%%%%%%%%%%%%%%%%%%%%%%%%%%
where $\tilde{\cal H}={\tilde a}^{\prime}/{\tilde a}$, $\tilde{\cal
F}={\tilde b}^{\prime}/{\tilde b}=-\tilde{\cal H}$
(conformal time is the same in both
frames). The conformal transformation (\ref{confor}) leads to
%%%%%%%%%%%%%%%%%%%%%%%%%%%%%%%%%%%%%%%%%%%%%%%%%%%%%%%%%%
\bq
 \widetilde a=\widetilde b^{-1}=(-\n)^{-1/(\sqrt{d+n}+1)}
\eq
%%%%%%%%%%%%%%%%%%%%%%%%%%%%%%%%%%%%%%%%%%%%%%%%%%%%%%%%%%
so that we obtain
%%%%%%%%%%%%%%%%%%%%%%%%%%%%%%%%%%%%%%%%%%%%%%%%%%%%%%%%%%
\bq
 h''_k+\frac{1}{\n} h'_k +k^2 h_k=0
\eq
%%%%%%%%%%%%%%%%%%%%%%%%%%%%%%%%%%%%%%%%%%%%%%%%%%%%%%%%%%
which implies, asymptotically, the same logarithmic mild growth
(\ref{gravsol}), as before. This is in complete agreement with the
frame-independence of the perturbation spectrum, already stressed in
\cite{GV2,GV3}.

\renewcommand{\theequation}{3.\arabic{equation}}
\setcounter{equation}{0}
\section{The growing mode of scalar perturbations}

We now turn to scalar perturbations, considering a four-dimensional,
conformally flat cosmological background
%%%%%%%%%%%%%%%%%%%%%%%%%%%%%%%%%%%%%%%%%%%%%%%%%%%%%%%%%%
\bq
 g_{\mu\nu}= a^2 diag (1, -\da_{ij}),\hspace{.5in} \vphi=\vphi(\n)
\label{homoback}\eq
%%%%%%%%%%%%%%%%%%%%%%%%%%%%%%%%%%%%%%%%%%%%%%%%%%%%%%%%%%
for which the Einstein equations can be written as
%%%%%%%%%%%%%%%%%%%%%%%%%%%%%%%%%%%%%%%%%%%%%%%%%
\brr
12\H^2&=&\vphi'^2\nonumber \\
8 \H'+4\H^2&=&-\vphi'^2\nonumber \\
\vphi''+2\H\vphi'&=&0
\label{background}
\err
%%%%%%%%%%%%%%%%%%%%%%%%%%%%%%%%%%%%%%%%%%%%%%%%%%%%%%%%%%
and the general expression for the (scalar part) of the perturbed
line element has
the well
known form (see for example \cite{MFB})
%%%%%%%%%%%%%%%%%%%%%%%%%%%%%%%%%%%%%%%%%%%%%%%%%%%%%%%%%%
\bq
 ds^2 =a^2 (1+2\phi) d\n^2-
a^2 \left[(1-2\psi) \d_{ij}+ 2\der_i\der_j E\right] dx^i dx^j
-2 a^2 \der_i B dx^i d\n
\label{sca}
\eq

A popular choice for discussing scalar perturbations of the metric
and
of the matter sources (in this case the dilaton field)
is the
so-called longitudinal gauge ($E=0=B$), where
$$
 \d g_{00} =2 a^2 \phi,~~~~\d g_{ij} =2 a^2 \psi\d_{ij},~~~~
\d g^{00}=-{2\over a^2}\phi,~~~~
\d g^{ij} =-\frac{2} {a^2} \psi\d^{ij}
$$
\bq
 \d g_{0i} = 0= \d g^{0i},~~~~~~~~~~~~~~~
  \d\vphi=\chi
\label{def}
\eq
%%%%%%%%%%%%%%%%%%%%%%%%%%%%%%%%%%%%%%%%%%%%%%%%%%%%%%%%%%
and where $\phi$ and $\psi$ turn out to coincide (to first order)
with the two gauge
invariant  Bardeen variables \cite{Bardeen,MFB}.
By perturbing Einstein's equations in this gauge,  we get,
from the off-diagonal spatial components, the condition $\phi=\psi$.
The remaining perturbation equations when written
explicitly in terms of
$\psi$ and $\chi$, take the form \cite{GV3,MFB}
%%%%%%%%%%%%%%%%%%%%%%%%%%%%%%%%%%%%%%%%%%%%%%%%%%%%%%%%%%
\bq
\nabla^2\psi-3\H \psi' =\frac{1}{4}\vphi'\chi'
\label{pert1}
\eq
%%%%%%%%%%%%%%%%%%%%%%%%%%%%%%%%%%%%%%%%%%%%%%%%%%%%%%%%%%
\bq
\psi''+3\H\psi'=\frac{1}{4}\vphi'\chi'
\label{pert2}
\eq
%%%%%%%%%%%%%%%%%%%%%%%%%%%%%%%%%%%%%%%%%%%%%%%%%%%%%%%%%%
\bq
\chi''+2\H\chi'-\nabla^2\chi =4 \vphi'\psi'
\label{pert3}
\eq
%%%%%%%%%%%%%%%%%%%%%%%%%%%%%%%%%%%%%%%%%%%%%%%%%%%%%%%%%%
with the additional constraint
%%%%%%%%%%%%%%%%%%%%%%%%%%%%%%%%%%%%%%%%%%%%%%%%%%%%%%%%%%
\bq
 \psi'+\H\psi=\frac{1}{4}\vphi'\chi
\label{constraint}
\eq
%%%%%%%%%%%%%%%%%%%%%%%%%%%%%%%%%%%%%%%%%%%%%%%%%%%%%%%%%%
Their combination gives the decoupled equation for the Fourier mode
$\psi_{k}$ ($\nabla ^2 \psi_k=-k^2\psi_k$)
%%%%%%%%%%%%%%%%%%%%%%%%%%%%%%%%%%%%%%%%%%%%%%%%%%%%%%%%%%
\bq
 \psi''_k+6\H\psi'_k+k^2 \psi_k=0 .
\label{psieq}
\eq
%%%%%%%%%%%%%%%%%%%%%%%%%%%%%%%%%%%%%%%%%%%%%%%%%%%%%%%%%%

The solution of equations
(\ref{background}) representing, in the S-frame, a dilaton-driven,
accelerated inflationary  background corresponds, in the
E-frame, to a growing dilaton field, and to an
 accelerated contraction. The asymptotic behaviour of such a
background
for $\eta\rightarrow0_{-}$
 can be parametrized in conformal
time  as \cite{GV3}
%%%%%%%%%%%%%%%%%%%%%%%%%%%%%%%%%%%%%%%%%%%%%%%%%%%%%%%%%%
\bq
 a(\n)=(-\n)^{1/2},\hspace{.5in} \vphi(\n)= -\sqrt{12}\ln a \; .
\label{solback}
\eq
%%%%%%%%%%%%%%%%%%%%%%%%%%%%%%%%%%%%%%%%%%%%%%%%%%%%%%%%%%
Eq.(\ref{psieq})
is solved asymptotically ($|k\eta|<<1$) by
%%%%%%%%%%%%%%%%%%%%%%%%%%%%%%
\bq
\psi_k= c_1(k)\ln|k\eta|+ c_2(k) \frac{1}{\eta^2}
\label{solpert}
\eq
%%%%%%%%%%%%%%%%%%%%%%%%%%%%%%%%%
showing that the mode $\psi_k$,
far from being frozen outside  the horizon,
grows in time like $\eta^{-2}$.

If we are interested, in particular, in the evolution of
primordial vacuum fluctuations, the correct normalization
of $\psi_{k}$ is to be fixed in terms of the variable $v$
satisfying canonical commutation relations. One has \cite{MFB} :
%%%%%%%%%%%%%%%%%%%%%%%%%%%%%%%%%%%%%%%%%%%%%%%%%%%%%%%%%%
\brr
 \psi_k &=& -\frac{\vphi'}{4 M_P k^2} \left(\frac{v_k}{a}\right)'
\nonumber \\
v&=& a(\chi+\frac{\vphi'}{\H} \psi)
\label{canonical}
\err
%%%%%%%%%%%%%%%%%%%%%%%%%%%%%%%%%%%%%%%%%%%%%%%%%%%%%%%%%%
where $v$, which has correct canonical dimensions
$[v_{k}]=[k]^{-1/2}$,
 satisfies the equation
%%%%%%%%%%%%%%%%%%%%%%%%%%%%%%%%%%%%%%%%%%%%%%%%%%%%%%%%%%
\bq
v''_k+\left(k^2-\frac{a''}{a}\right) v_k=0 \; .
\label{canscal}\eq
%%%%%%%%%%%%%%%%%%%%%%%%%%%%%%%%%%%%%%%%%%%%%%%%%%%%%%%%%%
Note that this equation is precisely the same
as eq.(\ref{ueq}) for tensor
perturbations. In the background (\ref{solback})
the exact solution of this equation, which
represents for $|k\eta|>>1$ a freely oscillating  positive frequency
mode
normalized to the  vacuum state at $\eta\rightarrow -\infty$, is
given
in
terms of the second-type Hankel function $H^{(2)}_{\nu}$ as
%%%%%%%%%%%%%%%%%%%%%%%%%%%%%%%%%%%%%%%%%%%%%%%%%%%%%%%%%%
\bq
v_k(\n)=\n^{1/2} H_0^{(2)}(|k\n|)_{\hbox{\large${ \longrightarrow
\atop
\n\rightarrow -\infty}$}}  \frac{1}{\sqrt{k}} e^{-i k \n}
\eq
%%%%%%%%%%%%%%%%%%%%%%%%%%%%%%%%%%%%%%%%%%%%%%%%%%%%%%%%%%
Far outside the horizon, $|k\eta|<<1$, one  obtains the
asymptotic
normalized expression
%%%%%%%%%%%%%%%%%%%%%%%%%%%%%%%%%%%%%%%%%%%%%%%%%%%%%%%%%%
\bq
\left|v_k(\n)\right|\simeq
\frac{a}{a_{\rm HC}}
\frac{|\ln (-k\n)|}{\sqrt{k}}
\label{asv}
\eq
%%%%%%%%%%%%%%%%%%%%%%%%%%%%%%%%%%%%%%%%%%%%%%%%%%%%%%%%%%
 which, when inserted into eq. (\ref{canonical}), yields the
following
expression for the typical amplitude of
fluctuations on length scales $k^{-1}$ at time $\eta$,
%%%%%%%%%%%%%%%%%%%%%%%%%%%%%%%%%%%%%%%%%%%%%%%%%%%%%%%%%%
\brr
    \left|\d _{\psi_k}(\n)\right| & = &
  k^{3/2} \left|\psi _{k}(\n)\right| \simeq
\frac{1}{ M_p |k\n|^2}
\left(\frac{k}{a}\right)_{\rm HC} \nonumber \\
&\simeq&\frac{H_1}{M_p} \frac{|k\n_1|^{3/2}}{ |k\n|^2}\; .
\err
%%%%%%%%%%%%%%%%%%%%%%%%%%%%%%%%%%%%%%%%%%%%%%%%%%%%%%%%%%
(to obtain the last equality we have multiplied and divided by the
final
inflationary scale
$H_{1}\simeq
(a_{1}\eta_{1})^{-1}$). In our background (\ref{solback}), the linear
approximation (i.e.
$  |\delta\psi_{k}|< 1$ ) is thus only valid on scales $k$  such that
%%%%%%%%%%%%%%%%%%%%%%%%%%%%%%%%%%%%%%%%%%%%%%%%%%%%%%%%%%
\bq
 \left|\frac{\n}{\n_1}\right|
{\ \lower-1.2pt\vbox{\hbox{\rlap{$>$}\lower5pt\vbox{\hbox{$\sim$}}}}
\ } \left(\frac{H_1}{M_p}\right)^{1/2}
\left(\frac{k_1}{k}\right)^{1/4}\; .
\label{range}
\eq
%%%%%%%%%%%%%%%%%%%%%%%%%%%%%%%%%%%%%%%%%%%%%%%%%%%%%%%%%%
As an example, for a nearly Planckian inflation scale $H_1$, this
condition implies that fluctuations over the scale presently probed
by COBE
observations can be treated perturbatively only for
$|\eta|>10^7\eta_1$ ($|t|> 10^{10}t_p$, in cosmic time). For a
similar result in a Kaluza-Klein context see ref. \cite{abbott}.

It is amusing to
observe that this conclusion can be evaded in the case of a
background
with $d>3$ isotropic spatial dimensions, in spite of the fact that
the
solution
is still growing in time. We refer to the Appendix for a detailed
discussion of
this case, as well as for a discussion of scalar perturbations in the
higher-dimensional anisotropic background (\ref{ddimback}).
The way out of this apparently disastrous result will be discussed in
the next two Sections.

\renewcommand{\theequation}{4.\arabic{equation}}
\setcounter{equation}{0}
\section{Covariant approach to scalar perturbations}

As discussed in the previous Section, the growth of scalar metric
perturbations outside  the horizon in the longitudinal gauge seems to
imply
that, in general, a linearized approach is not sufficient for a
complete and
consistent description of the evolution of scalar metric and dilaton
perturbations in
a dilaton-driven inflationary background. This is somewhat
surprising,
since on general grounds one would expect physical observables over
a given
scale to freeze out when such a scale goes outside the horizon. In
this Section, by using the fully covariant and gauge-invariant
formalism recently proposed in
\cite{BE}, we will show that the total energy density contained in
the
perturbations
is small compared to that of the background so that the physical
situation is  well described by a nearly homogeneous background.

Within the  covariant approach, one has to define two appropriate
variables,
$\Delta$ and $C$, characterizing the evolution of
density and
curvature inhomogeneities. For a scalar field dominated background,
such
variables are related, respectively,
to the comoving spatial Laplacian of the momentum density
$|\nabla_{\mu}
\vphi|$ of the scalar field, and to the comoving spatial Laplacian of
the
spatial part of the scalar curvature. Here ``spatial"  means
orthogonal
to the
direction of the  four vector $\nabla_{\mu}\vphi$, assumed to be
time-like, and
defining the
preferred world-lines of comoving observers. The exact definition of
these
variables for $d=3$, and in the absence of scalar field potential is
\cite{BE}
%%%%%%%%%%%%%%%%%%%%%%%%%%%%%%%%%%%%%%%%%%%%%%%%%%%%%%%%%%
\bq
\D= 2 a h_\mu^{\ \a}\nabla_\a\left(\frac{a}{f} h^{\mu\b}\nabla_{\b}
f\right)
\label{deltadef}
\eq
%%%%%%%%%%%%%%%%%%%%%%%%%%%%%%%%%%%%%%%%%%%%%%%%%%%%%%%%%%
%%%%%%%%%%%%%%%%%%%%%%%%%%%%%%%%%%%%%%%%%%%%%%%%%%%%%%%%%%
\bq
C= a h_\mu^{\ \a}\nabla_\a\left(a^3 h^{\mu\b}\nabla_{\b}\
{}^{(3)}R\right)
\label{cdef}
\eq
%%%%%%%%%%%%%%%%%%%%%%%%%%%%%%%%%%%%%%%%%%%%%%%%%%%%%%%%%%
where $f$  is the momentum density magnitude
%%%%%%%%%%%%%%%%%%%%%%%%%%%%%%%%%%%%%%%%%%%%%%%%%%%%%%%%%%
\bq
f=  \sqrt{\nabla_\mu \vphi\nabla^{\mu} \vphi},
\eq
%%%%%%%%%%%%%%%%%%%%%%%%%%%%%%%%%%%%%%%%%%%%%%%%%%%%%%%%%%
$h_{\mu\nu}$ is the projection tensor on the 3-space orthogonal to
the
momentum,
%%%%%%%%%%%%%%%%%%%%%%%%%%%%%%%%%%%%%%%%%%%%%%%%%%%%%%%%%%
\bq
h_{\mu\nu}=g_{\mu\nu}-u_\mu u_\nu,\hspace{.3in} u_\mu=-\frac{1}{f}
\nabla_\mu\vphi,
\eq
%%%%%%%%%%%%%%%%%%%%%%%%%%%%%%%%%%%%%%%%%%%%%%%%%%%%%%%%%%
and $^{(3)}R$ is the Ricci scalar of the spatial sub-manifold
orthogonal
to
$u^{\mu}$, defined in terms of the local expansion parameter
$\Theta$,
and the
shear
tensor $\sigma_{\mu\nu}$, as
%%%%%%%%%%%%%%%%%%%%%%%%%%%%%%%%%%%%%%%%%%%%%%%%%%%%%%%%%%
\bq
{}^{(3)}R= -\frac{2}{3} \T^2 +\s_{\mu\nu}\s^{\mu\nu}
+\hf \nabla_\mu\vphi \nabla^\mu \vphi
\eq
%%%%%%%%%%%%%%%%%%%%%%%%%%%%%%%%%%%%%%%%%%%%%%%%%%%%%%%%%%
%%%%%%%%%%%%%%%%%%%%%%%%%%%%%%%%%%%%%%%%%%%%%%%%%%%%%%%%%%
\bq
\T=-\nabla_\mu\left(\frac{1}{f} \nabla^\mu\vphi \right)\hspace{.8in}
\eq
%%%%%%%%%%%%%%%%%%%%%%%%%%%%%%%%%%%%%%%%%%%%%%%%%%%%%%%%%%
\bq
\s_{\mu\nu}=-\frac{1}{f}h_\mu^{\ \a}h_\nu^{\ \b}
\nabla_\a\nabla_\b\vphi
-\frac{1}{3} h_{\mu\nu}\T \; .
\eq
%%%%%%%%%%%%%%%%%%%%%%%%%%%%%%%%%%%%%%%%%%%%%%%%%%%%%%%%%%
The variables $\Da$ and $C$ satisfy exact equations which are
obtained
by taking the spatial gradient of the energy conservation and of the
Raychaudury equation \cite{EB}.

 From now on we will assume that the  unperturbed background is a
three-dimensional, spatially flat isotropic manifold described by the
solution
(\ref{solback}) of the string cosmology equations, and we shall
perform a
perturbative expansion around this background, in terms of the
variables
$\Delta$ and $C$.
One finds that, to zeroth order, the background values of $\Delta$
and
$C$ are
both vanishing, an obvious result for variables representing density
and
curvature fluctuations  when computed in a perfectly homogeneous
manifold. This
is, in fact, the main reason  why these variables  are
particularly suited for our  situation.

To the first order, by linearizing around the given background
the exact equations satisfied by  $\Delta$
and $C$, we find the following set of coupled first order equations
%%%%%%%%%%%%%%%%%%%%%%%%%%%%%%%%%%%%%%%%%%%%%%%%%%%%%%%%%%
\brr
2 \H\D'&=&C \nonumber \\
C'&=&2 \H \nabla^2 \D
\label{deltac}
\err
%%%%%%%%%%%%%%%%%%%%%%%%%%%%%%%%%%%%%%%%%%%%%%%%%%%%%%%%%%
which yields, upon differentiation,  a set of decoupled
second
order equations for the Fourier
modes $\Delta_{k}$, $C_{k}$
%%%%%%%%%%%%%%%%%%%%%%%%%%%%%%%%%%%%%%%%%%%%%%%%%%%%%%%%%%
\brr
\D''_k-2\H \D'_k+k^2 \D_k&=&0 \nonumber \\
C''_k+2\H C'_k+k^2 C_k&=&0
\label{secondord}
\err
%%%%%%%%%%%%%%%%%%%%%%%%%%%%%%%%%%%%%%%%%%%%%%%%%%%%%%%%%%
The general exact solution of these  equations can be written in
terms
of
Hankel functions of the first and second kind
%%%%%%%%%%%%%%%%%%%%%%%%%%%%%%%%%%%%%%%%%%%%%%%%%%%%%%%%%%
\brr
\D_k=&c_1 \n H_1^{(1)}(k\n) +c_2 \n H_1^{(2)}(k\n) \nonumber \\
C_k=&c_3  H_0^{(1)}(k\n) +c_4  H_0^{(2)}(k\n)\; .
\label{solutiondeltac}
\err
%%%%%%%%%%%%%%%%%%%%%%%%%%%%%%%%%%%%%%%%%%%%%%%%%%%%%%%%%%
The above solution  is consistent with the system (\ref{deltac})
provided
%%%%%%%%%%%%%%%%%%%%%%%%%%%%%%%%%%%%%%%%%%%%%%%%%%%%%%%%%%
\bq
c_3=k c_1,\hspace{.3in} c_4=k c_2 \; .
\label{conditions}
\eq
%%%%%%%%%%%%%%%%%%%%%%%%%%%%%%%%%%%%%%%%%%%%%%%%%%%%%%%%%%
{}From (\ref{solutiondeltac}) we  obtain
the
following asymptotic ($|k\eta|\rightarrow 0$)
behaviour for $\Delta$ and $C$
%%%%%%%%%%%%%%%%%%%%%%%%%%%%%%%%%%%%%%%%%%%%%%%%%%%%%%%%%%
\brr
\D_k&=&A_1(k) +\left[A_2(k)+A_3(k)\ln|k\n| \right]|k\n|^2 \nonumber
\\
C_k&=&B_1(k) +B_2(k) \ln |k\n|\; .
\label{solasint}
\err
%%%%%%%%%%%%%%%%%%%%%%%%%%%%%%%%%%%%%%%%%%%%%%%%%%%%%%%%%%
In this expression, only two of the coefficients
$A_{1,2,3}$, $B_{1,2}$ are arbitrary integration constants, while the
others
follow   from the condition  (\ref{conditions}) and the small
argument
limit of the  Hankel functions.

The fact that, in the linear approximation, $\Delta_{k}$ and $C_{k}$
stay
constant outside  the horizon, with at most the logarithmic variation
already found for
 tensor perturbations
(see Section 2), suggests that such variables
could
provide a consistent linearized description of the evolution of
vacuum fluctuations in terms of a perturbative expansion around a
homogeneous
background. To check this,  we   first have to
normalize
$\Delta_{k}$ and $C_{k}$ to the vacuum fluctuation  spectrum, by
relating them
to the canonical variable $v$ which defines the initial
vacuum state at $\eta\rightarrow-\infty$. This can always be   done,
for any   given mode $k$, by expressing $\Delta_{k}$ and $C_{k}$ to
first
order in terms of   the metric and dilaton scalar perturbation
variables, at
early enough time   scales, when the linear approximation is valid
also
in the
longitudinal gauge. Such a   relation  between linearized variables
can be
consistently established even for   modes
outside of the horizon, as discussed in the previous Section,
provided
the
corresponding time scale $\eta$ is in the interval (see eq.
(\ref{range}))
%%%%%%%%%%%%%%%%%%%%%%%%%%%%%%%%%%%%%%%%%%%%%%%%%%%%%%%%%%
\bq
\left(\frac{k_1}{k}\right)^{1/4}<\left|\frac{\n}{\n_1}\right|
<\frac{k_1}{k}
\label{goodint}
\eq
%%%%%%%%%%%%%%%%%%%%%%%%%%%%%%%%%%%%%%%%%%%%%%%%%%%%%%%%%%
(we have assumed $H_{1}{\
\lower-1.2pt\vbox{\hbox{\rlap{$<$}\lower5pt\vbox{\hbox{$\sim$}}}}\ }
M_{P}$).

By computing $\delta f$, $\delta h_{\mu\nu}$, and $\delta ^{(3)}R$ to
first
order in the scalar perturbations (\ref{def}), and using the
background
equations (\ref{background}), we obtain from the exact definitions of
$\Delta$
and $C$  (\ref{deltadef}), (\ref{cdef}) their explicit relation to
the
metric
perturbation variables, valid in the linear approximation,
%%%%%%%%%%%%%%%%%%%%%%%%%%%%%%%%%%%%%%%%%%%%%%%%%%%%%%%%%%
\bq
\D= 2 \nabla^2\left(\frac{3\H}{\vphi'}\chi-\psi+\frac{\chi'}{\vphi'}
\right)
\label{lindelta}
\eq
%%%%%%%%%%%%%%%%%%%%%%%%%%%%%%%%%%%%%%%%%%%%%%%%%%%%%%%%%%
%%%%%%%%%%%%%%%%%%%%%%%%%%%%%%%%%%%%%%%%%%%%%%%%%%%%%%%%%%
\bq
C= \frac{4\H}{\vphi'} \nabla^2\left(\nabla^2\chi+3\vphi'\psi'
+3\H \chi' \right)
\label{linc}
\eq
%%%%%%%%%%%%%%%%%%%%%%%%%%%%%%%%%%%%%%%%%%%%%%%%%%%%%%%%%%
These two relations have the remarkable feature that, while each term
on
the
right hand side  grows, asymptotically, as $1/\n^2$ or $1/\n^4$,  the
particular combinations entering in $\Delta$ and $C$ lead
  to an exact cancellation of the growing
mode contribution and reproduce the ``regularized"
asymptotic behaviour (\ref{solasint}). This cancellation can be
explicitly displayed
by noticing
that, using the background equations, the perturbation equations
(\ref{pert1})-(\ref{constraint}), and the definitions
(\ref{canonical}), the terms on the right hand side of eqs.
(\ref{lindelta}),
(\ref{linc})
can be combined to give
%%%%%%%%%%%%%%%%%%%%%%%%%%%%%%%%%%%%%%%%%%%%%%%%%%%%%%%%%%
\brr
\D&=& \frac{2}{\vphi'}\nabla^2\left(\frac{v}{a}\right)'
\nonumber\\
 C&=& \frac{4\H}{\vphi'} \nabla^2\nabla^2 \left(\frac{v}{a}\right)
\label{rel}
\err
%%%%%%%%%%%%%%%%%%%%%%%%%%%%%%%%%%%%%%%%%%%%%%%%%%%%%%%%%%
By inserting now the asymptotic solution (\ref{asv}) for the mode
$v_{k}$,
 we obtain for $\Delta_{k}$ and $C_{k}$ the normalized asymptotic
behaviour
%%%%%%%%%%%%%%%%%%%%%%%%%%%%%%%%%%%%%%%%%%%%%%%%%%%%%%%%%%
\brr
\left|\D_k\right|&\simeq& \frac{\sqrt{  k}}{M_p |a\n|_{\rm HC}}
\nonumber\\
 \left|C_k\right|&\simeq& \frac{ k^{5/2}}{M_p |a\n|_{\rm HC}} \ln
|k\n|
\err
%%%%%%%%%%%%%%%%%%%%%%%%%%%%%%%%%%%%%%%%%%%%%%%%%%%%%%%%%%
in full agreement with eq. (\ref{solasint}).

Once we have the normalized behaviour of $\Delta_{k}$ and $C_{k}$ we
can check
the validity of the linear approximation for such variables. The
typical
amplitudes of the vacuum fluctuations described by $\Delta$ and
$C$ over
length scales $k^{-1}$ can be estimated, respectively,
as $  k^{3/2}|\Delta_{k}|$ and $   k^{3/2}|C_{k}|$.
An approximate description of $\Delta$ and $C$  as small
perturbations
around
a homogeneous background is consistent provided their amplitude is
smaller
than the magnitude of the corresponding terms obtained by replacing
spatial
with temporal gradients in the exact definitions
(\ref{deltadef}),
(\ref{cdef}). Such terms are typically of order $\eta^{-2}$ for
$\Delta$ and
$\eta^{-4}$ for $C$. A linearized description of the evolution of the
vacuum
fluctuations in terms of $\Delta$ and $C$ is thus consistent if
%%%%%%%%%%%%%%%%%%%%%%%%%%%%%%%%%%%%%%%%%%%%%%%%%%%%%%%%%%
\brr
   \frac{ |k\n|^2}{M_p |a\n|_{\rm HC}}&\simeq&
\frac{H_1}{M_p}|k\n|^2|k\n_1|^{3/2}\ln|k\n|<1
\nonumber\\
   \frac{|k\n|^4}{M_p |a\n|_{\rm HC}}\ln|k\n|&\simeq&
\frac{H_1}{M_p}|k\n|^4|k\n_1|^{3/2}\ln|k\n|<1
\label{conditionsss}
\err
%%%%%%%%%%%%%%%%%%%%%%%%%%%%%%%%%%%%%%%%%%%%%%%%%%%%%%%%%%
Both conditions are clearly satisfied, at all $|\eta|\ge \eta_{1}$,
for
all
$k\le 1/\n$, provided $H_{1}{\
\lower-1.2pt\vbox{\hbox{\rlap{$<$}\lower5pt\vbox{\hbox{$\sim$}}}}\ }
M_{P}$.

As a consistency check we can easily verify that the spectral
amplitude
of the
fluctuations $\delta\rho/\rho$ of the comoving source energy density,
defined
in the linear approximation as
%%%%%%%%%%%%%%%%%%%%%%%%%%%%%%%%%%%%%%%%%%%%%%%%%%%%%%%%%%
\bq
  k^{3/2}\left|\frac{\d\rho_k}{\rho}\right|\simeq
\frac{1}{\sqrt{k}} |\D_k|\simeq \frac{H_1}{M_p}|k\n_1|^{3/2} \ln|k\n|
\label{spectralamplitude}
\eq
%%%%%%%%%%%%%%%%%%%%%%%%%%%%%%%%%%%%%%%%%%%%%%%%%%%%%%%%%%
is smaller than  critical for any mode $k\le k_{1}=\eta_{1}^{-1}$.
This justifies the fact that the evolution of the vacuum fluctuations
is
treated linearly, neglecting their back-reaction on the original
geometry.

Note that the previous equation defines
fluctuations of the total {\it comoving} energy density, where
comoving
is
referred to the time-like momentum of the scalar field, i.e.
%%%%%%%%%%%%%%%%%%%%%%%%%%%%%%%%%%%%%%%%%%%%%%%%%%%%%%%%%%
\bq
\rho = T_{\mu\nu} u^\mu u^\nu=
 \left(
\der_\mu\vphi\der_\nu\vphi -\hf g_{\mu\nu}
\der_\a\vphi\der^\a\vphi\right)
\frac{\der^\mu\vphi\der^\nu\vphi}{\der_\a\vphi\der^\a\vphi}
\label{tmunu}
\eq
%%%%%%%%%%%%%%%%%%%%%%%%%%%%%%%%%%%%%%%%%%%%%%%%%%%%%%%%%%
Eq. (\ref{tmunu}) contains contributions from both metric and dilaton
perturbations, as defined in the longitudinal gauge. In
the covariant approach that we are considering, each one of the two
contributions
cannot be separately computed as it would turn out to be too
large to
be consistent with a linearized treatment. Only the appropriate
combination
corresponding to $\Delta$  remains small enough to be treated
perturbatively.

Moreover,   the spectral distribution
(\ref{spectralamplitude}) is the same as the
one  obtained for $\psi$ and $\chi$ separately, if the growing
solution
of the perturbation equations is simply neglected \cite{GV3}. It also
corresponds to the dilaton and graviton spectrum obtained via a
Bogoliubov
transformation, connecting the initial vacuum to the final vacuum
state
of a radiation-dominated background
(i.e. to the spectrum defined with
respect to the asymptotic particle content of the amplified
fluctuations
\cite{GV3}). This coincidence suggests that  the
growing mode
of scalar  metric perturbations  does not have  any direct  physical
meaning. If so it should be possible to get rid of it
through a suitable coordinate choice.
 This possibility will be discussed in the next section.

As far as the
fluctuations in
the ``geometric" (scalar curvature) part of the energy
density are concerned, the spectral amplitude obtained from
$C_{k}$,
%%%%%%%%%%%%%%%%%%%%%%%%%%%%%%%%%%%%%%%%%%%%%%%%%%%%%%%%%%
\bq
\frac{1}{k^{5/2}}|C_k|\simeq
\frac{H_1}{M_p}|k\n_1|^{3/2}\ln|k\n|
\label{spectralamplitude2}
\eq
%%%%%%%%%%%%%%%%%%%%%%%%%%%%%%%%%%%%%%%%%%%%%%%%%%%%%%%%%%
exactly reproduces (\ref{spectralamplitude}) (even in the logarithm!)
and
also coincides with the spectral behaviour of tensor
perturbations (see eq. (\ref{4damplitude})).
In the case of the $C_{k}$ spectrum we are also not dealing with a
purely
gravitational energy distribution: again the
contributions of metric and dilaton fluctuations, as defined in the
longitudinal gauge, are both present and mixed.

We note, finally, that in the linear approximation the fluctuations
in
both
energy density and curvature are defined in terms of the canonical
variable
$v$. Indeed, from eq. (\ref{rel}), (\ref{spectralamplitude}) and
the
exact definition
(\ref{cdef}), we find
%%%%%%%%%%%%%%%%%%%%%%%%%%%%%%%%%%%%%%%%%%%%%%%%%%%%%%%%%%
\brr
\frac{\d \rho}{\rho} &\simeq& \frac{v}{a}
\nonumber\\  \d\ {}^{(3)}R &\simeq&\nabla^2 \frac{v}{a}
\err
%%%%%%%%%%%%%%%%%%%%%%%%%%%%%%%%%%%%%%%%%%%%%%%%%%%%%%%%%%
The fact that $\D$ and $C$  are small may thus be
seen to follow from the fact that the contributions of $\psi$ and
$\chi$ combine just  to give $v$. It is
striking  that the linearized asymptotic behaviour (\ref{solasint})
of $\Delta$ and $C$
 is correctly given  by
extrapolating the logarithmic behaviour of $v$, eq. (\ref{asv}),
to times at which
the definition of $v$ in terms of $\psi$ and $\chi$ is no longer
consistent with the linear  perturbation theory. This suggests
that $v$, first identified in \cite{MU} as the correct variable for
the
canonical
quantization of perturbations in the linear approximation, could
also be an
appropriate variable for a consistent perturbative expansion
describing the
evolution of inhomogeneities in a general scalar-tensor background.

\renewcommand{\theequation}{5.\arabic{equation}}
\setcounter{equation}{0}
\section{Gauging down the growing mode}

The results of the previous Section strongly suggest
that, owing to the special properties of the growing mode,
 physically observable inhomogeneities stay   small at all  times.
 If so, a suitable gauge  reflecting that fact should exist.
Stated differently,  there should
be a good coordinate system in which the growing mode of scalar
 perturbations should be  strongly suppressed and metric
perturbations
 can be treated perturbatively throughout their evolution.
In this Section
we will present a suitable candidate for such a job, the
``off-diagonal"
gauge. As it
turns out, this choice  ``gauges away" completely the growing
mode at $k=0$ and ``gauges down" the small-$k$ growing modes.
Although this will be shown to be sufficient in order
to construct a perturbative solution around a ``shifted" background,
 we will argue that the construction of a systematic  expansion
in a small parameter may  require a further change of coordinates.

We start our search for the desired gauge from the longitudinal
gauge (\ref{def}), in which the line element takes the form
%%%%%%%%%%%%%%%%%%%%%%%%%%%%%%%%%%%%%%%%%%%%%%%%%%%%%%%%%%
\bq
ds^2 = a^2(\eta) [(1+2\phi) d\eta^2 - (1-2\psi)\da_{ij} dx^i dx^j] \;
{}.
\eq
%%%%%%%%%%%%%%%%%%%%%%%%%%%%%%%%%%%%%%%%%%%%%%%%%%%%%%%%%%
 Consider now the coordinate transformation
%%%%%%%%%%%%%%%%%%%%%%%%%%%%%%%%%%%%%%%%%%%%%%%%%%%%%%%%%%
\bq
\eta \rightarrow \tilde{\eta} = \eta + \theta(\eta, x^i) \;\;, \;
\tilde{x}^i = x^i \;
\eq
%%%%%%%%%%%%%%%%%%%%%%%%%%%%%%%%%%%%%%%%%%%%%%%%%%%%%%%%%%
 It is easy to check that, at first order in
$\theta, \phi, \psi$, the choice
%%%%%%%%%%%%%%%%%%%%%%%%%%%%%%%%%%%%%%%%%%%%%%%%%%%%%%%%%%
\bq
\theta=- \H^{-1} \psi
\label{fpsi}
\eq
%%%%%%%%%%%%%%%%%%%%%%%%%%%%%%%%%%%%%%%%%%%%%%%%%%%%%%%%%%
brings the line element to the ``off-diagonal" form
%%%%%%%%%%%%%%%%%%%%%%%%%%%%%%%%%%%%%%%%%%%%%%%%%%%%%%%%%%
\bq
ds^2 = a^2(\tilde{\eta}) \Biggl[\left(1 - 2 \theta' - 4\H \theta
+ 2(\phi -
\psi)\right)
d\tilde\eta^2
- 2 \partial_i \theta dx^i d\tilde\eta -\da_{ij} dx^i dx^i\Biggr]
\label{gaugedmetric}
\eq
%%%%%%%%%%%%%%%%%%%%%%%%%%%%%%%%%%%%%%%%%%%%%%%%%%%%%%%%%%
Furthermore,  at  first order, the relation $\phi = \psi$ holds true
for
the general solution while,
  for the growing mode solution, one also finds $\psi'= -4\H \psi$.
Using (\ref{fpsi}), this implies $\theta' = -2\H \theta$.
Therefore, as far as the growing mode is concerned,
 the line element (\ref{gaugedmetric}) simply becomes
%%%%%%%%%%%%%%%%%%%%%%%%%%%%%%%%%%%%%%%%%%%%%%%%%%%%%%%%%%
\bq
ds^2 = a^2(\tilde{\eta}) [(1+ \O(\tilde\eta^2 \partial^2
\phi))d\tilde\eta^2 -(\da_{ij}+\O(\tilde\eta^2 \partial_i \partial_j
\phi) dx^i
dx^i -2 \partial_i \theta
dx^i
d\tilde{\eta} ]
\eq
%%%%%%%%%%%%%%%%%%%%%%%%%%%%%%%%%%%%%%%%%%%%%%%%%%%%%%%%%%
We see that the dangerously large entries in $\d g_{00}$ and $\d
g_{ii}$
have been tamed and have given rise to the  off-diagonal  entry
 $\partial_i \theta  =- \H^{-1} \partial_i \psi$. For long
wavelengths
the typical size of the off-diagonal entry, $\d g_{0i}\sim |k\n|
\psi$,
 is a factor $|k\n|<<1$ smaller
than the original perturbation. Even smaller terms appear in $\d
g_{00}$,
 $\d g_{ii}$ and are such that the growing mode is completely
gauged-away
for an exactly homogeneous perturbation.

The previous result suggests starting the analysis of scalar
perturbations
directly in a new  ``off-diagonal" gauge defined,
according to the general line element (\ref{sca}), by
%%%%%%%%%%%%%%%%%%%%%%%%%%%%%%%%%%%%%%%%%%%%%%%%%%%%%%%%%%
\bq
ds^2 = a^2(\eta) [(1+2\phi) d\eta^2 - \da_{ij} dx^i dx^i-
2 \der_i B dx^i d\eta]
\eq
%%%%%%%%%%%%%%%%%%%%%%%%%%%%%%%%%%%%%%%%%%%%%%%%%%%%%%%%%%
{}from which the metric perturbations can be read
%%%%%%%%%%%%%%%%%%%%%%%%%%%%%%%%%%%%%%%%%%%%%%%%%%%%%%%%%%
\brr
 \d g_{00} &=&2 a^2 \phi \nonumber \\
 \d g_{i0} &=& - a^2 \der_i B\nonumber \\
\d g_{ij} &=&0
\label{nondiagonal}
\err
%%%%%%%%%%%%%%%%%%%%%%%%%%%%%%%%%%%%%%%%%%%%%%%%%%%%%%%%%%
and, to first order,
\brr
  \d g^{00} &=&- 2 \phi /a^2 \nonumber \\
  \d g^{i0} &=&- \der_i B /a^2 \nonumber \\
\d g^{ij} &=&0 \; .
\label{nondiagonalinv}
\err
This gauge choice
is interesting in itself. It represents  a complete
gauge
choice, namely, it does not contain any residual degrees of freedom
and
it is
similar, in that respect, to the longitudinal gauge. Indeed, under
an
infinitesimal
coordinate transformation which preserve the scalar nature of the
perturbations, $x^{\mu}\rightarrow\tilde
x^{\mu}=
x^{\mu} + \epsilon^{\mu}(x^{\alpha})$, with
$\epsilon^{0}=\epsilon^{0}(\eta,
\vec x)$, $\epsilon^{i}=\partial^{i} \epsilon(\eta,\vec x)$, the
various
entries of the general perturbed metric (\ref{sca})
transform as follows \cite{MFB}
%%%%%%%%%%%%%%%%%%%%%%%%%%%%%%%%%%%%%%%%%%%%%%%%%%%%%%%%%%
\brr
\phi \ra
\widetilde\phi &=&\phi-\frac{a'}{a} \epsilon^0-{\epsilon^0}
'\nonumber
\\
\psi \ra\widetilde\psi &=&\psi+\frac{a'}{a} \epsilon^0 \nonumber \\
E \ra  \widetilde E &=&E- \epsilon \nonumber \\
B \ra \widetilde B &=&B+ \epsilon^0-{\epsilon} '
\label{unchoice}
\err
%%%%%%%%%%%%%%%%%%%%%%%%%%%%%%%%%%%%%%%%%%%%%%%%%%%%%%%%%%
so that the choice  $\tilde E=0 = \tilde\psi$ indeed determines the
vector
$\epsilon^{\mu}$ uniquely.

By perturbing,  in this gauge,  the Einstein equations around the
background (\ref{background}), we obtain from the $(0,0)$
and
$(i,i)$ components of eq.(\ref{einstein}),  respectively, the
perturbation
equations
%%%%%%%%%%%%%%%%%%%%%%%%%%%%%%%%%%%%%%%%%%%%%%%%%%%%%%%%%%
\bq
 -4 \H \nabla^2 B
=\chi'\vphi'
\label{nonlongpert1}
\eq
%%%%%%%%%%%%%%%%%%%%%%%%%%%%%%%%%%%%%%%%%%%%%%%%%%%%%%%%%%
%%%%%%%%%%%%%%%%%%%%%%%%%%%%%%%%%%%%%%%%%%%%%%%%%%%%%%%%%%
\bq
4 \H \phi'
=\chi'\vphi' \; ,
\label{nonlongpert2}
\eq
%%%%%%%%%%%%%%%%%%%%%%%%%%%%%%%%%%%%%%%%%%%%%%%%%%%%%%%%%%
from which the simple relation $\phi' = - \nabla^2 B$ also follows.
{}From the $(i,0)$ and $(i,j\ne i)$  components we obtain,
respectively,
two constraints expressing $\chi$ and $\phi$ in terms of $B$,
%%%%%%%%%%%%%%%%%%%%%%%%%%%%%%%%%%%%%%%%%%%%%%%%%%%%%%%%%%
%%%%%%%%%%%%%%%%%%%%%%%%%%%%%%%%%%%%%%%%%%%%%%%%%%%%%%%%%%
\bq
 4 \H \phi  =\chi \vphi'
\label{nonlongpert3}
\eq
%%%%%%%%%%%%%%%%%%%%%%%%%%%%%%%%%%%%%%%%%%%%%%%%%%%%%%%%%%

%%%%%%%%%%%%%%%%%%%%%%%%%%%%%%%%%%%%%%%%%%%%%%%%%%%%%%%%%%
\bq
\phi = -(B'+2 \H B)
\label{nonlongpert4}
\eq
%%%%%%%%%%%%%%%%%%%%%%%%%%%%%%%%%%%%%%%%%%%%%%%%%%%%%%%%%%
{}From equations (\ref{nonlongpert1})-(\ref{nonlongpert4}) it is
possible to
obtain the following decoupled equation for
the Fourier mode $B_{k}$,
%%%%%%%%%%%%%%%%%%%%%%%%%%%%%%%%%%%%%%%%%%%%%%%%%%%%%%%%%%
\bq
B''_k +2 \H B'_k +(k^2-4 \H^2) B_k=0
\eq
%%%%%%%%%%%%%%%%%%%%%%%%%%%%%%%%%%%%%%%%%%%%%%%%%%%%%%%%%%
which, asymptotically  ($|k\eta|<< 1$), has the  solution
%%%%%%%%%%%%%%%%%%%%%%%%%%%%%%%%%%%%%%%%%%%%%%%%%%%%%%%%%%
\bq
B_k=c_1(k) \n \ln|k\eta|  + c_2(k) \n^{-1}
\label{soluzione}
\eq
%%%%%%%%%%%%%%%%%%%%%%%%%%%%%%%%%%%%%%%%%%%%%%%%%%%%%%%%%%

One can  check that $-\H B=-B/2 \eta$ has the same
asymptotic behaviour as $\psi$ in the longitudinal gauge, i.e.,
$\psi \sim
 \eta^{-2}$. This has to be the case since, in the off-diagonal
gauge, $-\H B$ corresponds exactly to the Bardeen variable
$\Psi$ which, instead, coincides with $\psi$
in the longitudinal gauge \cite{MFB}. Indeed, adding momentarily a
tilde
to quantities in the off-diagonal gauge, and using the asymptotic
solutions (\ref{soluzione}) and (\ref{solpert}) (with an obvious
rescaling of the integration constants)
%%%%%%%%%%%%%%%%%%%%%%%%%%%%%%%%%%%%%%%%%%%%%%%%%%%%%%%%%%
\bq
{\tilde \Psi}= {\tilde \psi} - {a' \over a}{\tilde B}=
c_1 \ln |k\eta|+{c_2\over \eta^2}=\psi= \Psi
\eq
%%%%%%%%%%%%%%%%%%%%%%%%%%%%%%%%%%%%%%%%%%%%%%%%%%%%%%%%%%
On the other hand, in the off-diagonal gauge
(\ref{nondiagonal}), the growing mode
solution does not contribute to the $\da g_{00}$ and $\da \vphi$
perturbations.
Inserting the solution (\ref{soluzione}) into
eqs.(\ref{nonlongpert3}), (\ref{nonlongpert4})
we find the harmless asymptotic behaviour
$\phi \simeq \chi \simeq const \times \ln |k\eta|$.

Another interesting property of the off-diagonal gauge is that the
canonical variable $v$, instead of having the complicated form of
eq.(\ref{canonical}), is just given by
 $a\chi$. As  a consequence, $\chi$, like $h$ in eq. (\ref{grav}),
 obeys the simple, decoupled equation
%%%%%%%%%%%%%%%%%%%%%%%%%%%%%%%%%%%%%%%%%%%%%%%%%%%%%%%%%%
\bq
\chi''+2\H \chi'-\nabla^2 \chi=0\; .
\eq
%%%%%%%%%%%%%%%%%%%%%%%%%%%%%%%%%%%%%%%%%%%%%%%%%%%%%%%%%%
whose asymptotic  solution has already been given in eq.
(\ref{GRAVSOL}).

Because of the extra power of $|k\n|$  in the
Fourier transform of $\da g_{i0}$, the off-diagonal gauge stands a
better chance of   providing  a setup for a reliable linear
description
 of  amplified vacuum fluctuations --both for the metric and
for the dilaton-- when they are far outside the horizon.
In order to verify
 this, we normalize the mode $B_k$ to the
initial vacuum state by using the  asymptotic relation
$ B =-2\eta \psi$ (Cf. eq.(\ref{fpsi})).
After inserting the correct normalization of $\psi_k$, we
obtain, for the   typical
amplitude of the fluctuations associated with
$\nabla B$, over length scales $k^{-1}$,
%%%%%%%%%%%%%%%%%%%%%%%%%%%%%%%%%%%%%%%%%%%%%%%%%%%%%%%%%%
\bq
|\da_{B_k}(\eta)|= k^{3/2}|k  B_k| \simeq {H_1\over
M_p}
|k\eta_1|^{1/2}\left|\eta_1 \over \eta \right|\; ,
\label{fiveb}
\eq
%%%%%%%%%%%%%%%%%%%%%%%%%%%%%%%%%%%%%%%%%%%%%%%%%%%%%%%%%%
which for any $k$ is smaller than $1$ for all $|\eta|>\eta_1$, namely
for the whole duration of the inflationary epoch. An even smaller
expression is easily obtained for $|\da_{{\phi}_k}(\eta)|$,
%%%%%%%%%%%%%%%%%%%%%%%%%%%%%%%%%%%%%%%%%%%%%%%%%%%%%%%%%%
\bq
 \left|\da_{{\phi}_k}(\eta)\right|\simeq
\left(\frac{H_1}{M_p}\right) (k\n_1)^{3/2}\ln|k\n| \; .
\eq
%%%%%%%%%%%%%%%%%%%%%%%%%%%%%%%%%%%%%%%%%%%%%%%%%%%%%%%%%%

A second significant check that amplified  vacuum
fluctuations do not perturb the homogeneous background in any
 substantial way follows  from an explicit
computation of the invariant
%%%%%%%%%%%%%%%%%%%%%%%%%%%%%%%%%%%%%%%%%%%%%%%%%%%%%%%%%%
\bq
W={|C_{\mu\nu\a\b}C^{\mu\nu\a\b}|\over |R_{\mu\nu}R^{\mu\nu}|}
\label{weyl}
\eq
%%%%%%%%%%%%%%%%%%%%%%%%%%%%%%%%%%%%%%%%%%%%%%%%%%%%%%%%%%
where $C_{\mu\nu\a\b}$
and $R_{\mu\nu}$ are, respectively, the Weyl and
Ricci
tensors. The unperturbed background is conformally flat and has a
vanishing
Weyl tensor. Consequently, the invariant  $W$  vanishes to zeroth
order
and to first order in metric perturbations.  To  second order in
metric perturbations, a straightforward but rather
long calculation (which we shall not reproduce here) shows that
 an upper bound on the magnitude of $W$ is given by  $W<|\pa_iB|^2$.
The magnitude of $W$ is thus bounded  by the fluctuation
$|\da_B|^2$,
which, according to
eq.(\ref{fiveb}),
remains smaller than unity on all scales and at
 all times $|\eta|>\eta_1$. This
result  represents an additional  covariant
and gauge-invariant confirmation that the physical manifold can
 be consistently described, to leading
order, in terms of some small inhomogeneity perturbations lying on
top
 of a homogeneous background solution of the
string cosmology equations.

A complete check of the validity of the linear
approximation of
cosmological  perturbation  expansion  requires a full understanding
of
the
formal structure of cosmological perturbation theory beyond leading
order, which, unfortunately, is lacking to this date.
An important step in this direction, which, to the
best of
our knowledge, has not been attempted before in any other approach,
 would
consist of a
direct comparison between second and first  order terms in the
perturbed
Einstein equations. We have undertaken part of such
 calculation by computing
all
quadratic terms in the four independent scalar perturbation
equations.
The
outcome will be discussed below,  not before
warning
the reader that a full second order computation of scalar
perturbations
should require  a generalization of the off-diagonal gauge
ansatz for the
metric, as
well as consideration of the mixing of scalar, tensor and possibly
vector
perturbations at second order.

We consider, for computational convenience,  the following
equivalent form of  Einstein's equations (\ref{einstein}),
%%%%%%%%%%%%%%%%%%%%%%%%%%%%%%%%%%%%%%%%%%%%%%%%%%%%%%%%%%
\bq
R_{\mu \nu} - \hf \der_\mu\vphi\der_\nu\vphi = 0
\eq
%%%%%%%%%%%%%%%%%%%%%%%%%%%%%%%%%%%%%%%%%%%%%%%%%%%%%%%%%%
whose left hand side we denote, for simplicity,  by $E_{\mu \nu}$.
For completeness we first write down  the first order
expressions for $E_{\mu \nu}$
which we denote by $E_{\mu \nu}^{(1)}$
%%%%%%%%%%%%%%%%%%%%%%%%%%%%%%%%%%%%%%%%%%%%%%%%%%%%%%%%%%
\brr
E_{00}^{(1)}&=& \nabla ^2(B'+\H B+\phi)+3\H \phi'-\vphi' \chi'
\nonumber \\
E_{0i}^{(1)}&=& 2 \H \phi_i -{1\over 2}\vphi' \chi_i
\nonumber \\
E_{ij}^{(1)}&=& -(B_{ij}'+2\H B_{ij}+\phi_{ij})-\H \da_{ij}(\phi'+
\nabla^2 B) \; ,
\label{maurizio1}
\err
%%%%%%%%%%%%%%%%%%%%%%%%%%%%%%%%%%%%%%%%%%%%%%%%%%%%%%%%%%
where subscripts on $B$, $\phi$ and $\chi$ denote spatial gradients.
The
solution of the first order equations
 has already been given in equations
(\ref{nonlongpert3}),  (\ref{nonlongpert4}) and (\ref{soluzione}).
Straightforward  but lengthy calculations lead to the following
quadratic expressions for the four independent components of
$E_{\mu \nu}$, which we denote by $\tilde{E}_{\mu \nu}^{(2)}$
(the full second order $E_{\mu \nu}^{(2)}$ includes, of course,
terms linear in the second order corrections to the fluctuations)
%%%%%%%%%%%%%%%%%%%%%%%%%%%%%%%%%%%%%%%%%%%%%%%%%%%%%%%%%%
\brr
E_{0 0}^{(2)} &=&  3\H B'_kB^k -\phi'\nabla ^2B -\phi_i^2+\H
\phi_kB^k
-6\H \phi \phi'- {1\over 2}(\chi ')^2 \nonumber \\
E_{0 i}^{(2)} &=&-\H B_i(\nabla ^2B +\phi') - B_k B'_{ik} -
B^k\phi_{ik} -\phi_i \nabla^2B- \nonumber \\ &-& 4\H \phi \phi_i +
\phi^k B_{ik}
- {1\over 2}\chi '  \chi_i \nonumber \\
E_{i j}^{(2)} &=&(\phi' +\nabla^2 B) B_{ij} -\H \da_{ij} B^kB_k'
-B_i^k B_{jk} +2\phi B_{ij}'+ \nonumber \\
&+&\H \da_{ij}(\phi_kB^k+4 \phi \phi' +
2\phi \nabla^2 B) +\phi_i\phi_j+2\phi \phi_{ij}+4\H \phi
B_{ij}-{1\over
2}
\chi_i \chi_j
\label{maurizio2}
\err
%%%%%%%%%%%%%%%%%%%%%%%%%%%%%%%%%%%%%%%%%%%%%%%%%%%%%%%%%%
We could now write down the perturbation equations up to
second order and try to solve them explicitly.
Since, for the time being, we
are only interested   in an order of magnitude estimate
of the
second order corrections  to the first order solution, we
will rather only keep track  of anomalously large terms
showing that, after an additional coordinate transformation,
they all vanish eventually.

 Large terms only originate from the growing mode in $B$.
Keeping track of those
 and using the first order equations
to simplify
the
second order equations, we arrive at the following structure
for the second order corrections to the metric fluctuations
\brr
 \d g_{00} &=&
2 a^2 [\phi - \hf B_i B_i + \O(\phi^2)] \nonumber \\
 \d g_{i0} &=& - a^2 [\der_i B +\O( \phi B_i)] \nonumber \\
\d g_{ij} &=& a^2 \O(\delta_{ij} \phi^2)
\label{nondiagonalimpa}
\err
It is quite remarkable that all the large terms in (\ref{maurizio2})
can be cancelled by the single large term $- \hf B_i B_i$ appearing
in $\d g_{00}$. At first sight this appears to indicate
that the perturbative expansion is breaking
down, since $ B_i B_i$, when evaluated using the growing mode of $B$,
is much larger than $\phi$. Consider, however, the
 coordinate transformation
\bq
\eta \ra \tilde{\eta} =\eta, ~~~~~~~~~~
x^i \ra \tilde{x}^i = x^i +\int^{\eta} B_i(\eta') d\eta'
\label{miraclet}\eq
As  easily verified, this  transformation
 eliminates simultaneously both the contribution of the growing
mode of
$B$ to the  off-diagonal ($d \eta , d x^i$)  entry and the nasty
$-\hf
B_i B_i$
term in $\da g_{00}$,
leaving only small corrections to the linear $\phi$ term
as well as  other small non-diagonal  terms, typically containing two
spatial derivatives acting on $B$. This cancellation is
highly  non-trivial
and  depends crucially on the value $- \hf$ of the coefficient of
$B_iB^i$ at second order in $\da g_{00}$ (see
eq.(\ref{nondiagonalimpa})).
After performing the coordinate transformation
(\ref{miraclet}), the new metric perturbations
(indicated by a tilde) up to
second order have the following structure
\brr
 \d g_{00} &=&
2 a^2 [\phi+\O(\phi^2)] \nonumber \\
 \d g_{i0} &=& - a^2 [ \O( \phi B_i)]\nonumber \\
\d g_{ij} &=&2a^2[ \int^{\eta} B_{ij} d\eta' - \hf \int^{\eta}
B_{ik}(\eta') d\eta' \int^{\eta} B_{jk}(\eta') d\eta'
+ \O( \d _{ij} \phi^2)]
\label{nondiagonalimp}
\err
We conclude that, after suitable
coordinate transformations, all second-order corrections are down by
at
least
an extra factor $\phi$ (or $ \int d\eta B_{ij} \sim \phi$)
relative to first
order, i.e. are genuinely small.

The above second order calculation provides  additional  support to
the
conclusion  that a linearized description of scalar perturbations is
generically adequate in an appropriate gauge.
 Of course, the calculation should not be interpreted as
suggesting
that perturbation theory will be uniformly  good at all scales. At
second order
different modes (as well as different
angular momenta) couple in a non-trivial way  and it is not excluded
that regions
of the spectrum which are very depressed at first order  will get
larger
contributions from second order corrections.
Investigating  whether this phenomenon
could enhance the power spectrum at very small $k$ is left
 as an interesting
subject for future research.

\renewcommand{\theequation}{6.\arabic{equation}}
\setcounter{equation}{0}
\section{Summary and conclusions}

As mentioned already in the introduction,  the cosmological equations
obtained from the low energy
string effective action imply that an arbitrarily small,
finite density of string matter is enough to trigger the evolution
of the  perturbative string
vacuum (taken as initial state)  towards a regime of
growing string coupling and curvature. Such a regime eventually
evolves
 into a long
period of dilaton-driven inflation \cite{GV3,GAS}. This final epoch
corresponds to a phase of accelerated expansion in the S-frame (of
accelerated contraction in the E-frame) and  is invariantly
characterized by shrinking event horizons \cite{GV1}-\cite{GV3}, in
contrast with the constant event horizon of the more conventional de
Sitter-like inflation. In such a context, the external metric and
dilaton background fields contribute jointly to
the parametric amplification of metric perturbations \cite{GG1}.

In this paper we have considered and contrasted tensor and scalar
metric
perturbations showing that, while for the former a straightforward
computation is possible, for the latter some special care is needed
owing to the presence of rapidly growing modes in the most
conventional
parametrization of the metric fluctuations (longitudinal gauge).
Nonetheless, we have been able to
compute the
scalar perturbation spectrum using either an appropriate
set of covariant and gauge
invariant
variables \cite{BE}, or by  using a new gauge, which we found to be
particularly useful for the purpose of keeping scalar metric
 perturbations small.

In spite of their very different treatment, tensor and scalar
perturbations
are predicted to have very similar amplitudes and spectra, given by
(see
eqs. (\ref{4damplitude}) and (\ref{spectralamplitude})):
\bq
\left|\d _{h_k}(\n)\right| \simeq
\left(\frac{H_1}{M_p}\right) (k\n_1)^{3/2}\ln|k\n|
\eq
and
\bq
  k^{3/2}\left|\frac{\d\rho_k}{\rho}\right|\simeq
\frac{1}{\sqrt{k}} |\D_k|\simeq \frac{H_1}{M_p}|k\n_1|^{3/2} \ln|k\n|
\eq

These perturbations should manifest themselves in (at least) two
different ways:
on one hand, as metric perturbations at the surface of last
scattering, they will affect the homogeneity of the CMB spectrum
(through
 the Sachs-Wolfe effect). As discussed elsewhere \cite{GV1}, such an
effect will be small at the scales measured by COBE, even if one takes
$H_1\sim M_p$. This is simply because
our spectrum, unlike the one of (almost) de-Sitter inflation, is
decreasing with the length-scale of the perturbation.

The second effect is a background of relic gravitons and dilatons
which should be still around us having been left over from the
``Planck-String" era. Quite amusingly,
both spectra bear a strong resemblance to the (unperturbed) Planckian
spectrum of the CMB photons themselves.
Indeed, the above equations readily lead to
the graviton-dilaton spectral-energy distribution:
\bq
\frac{d \rho}{d \ln \omega} \simeq \omega^4
\left(\frac{\omega_1}{\omega}\right)
\ln^2 \left(\frac{\omega_1}{\omega}\right) \; , \label{spectra}
\eq
at $\omega < \omega_1$ (and exponentially suppressed
 at $\omega > \omega_1$). Here $\omega_1$
is (the present value of) the maximal amplified (red-shifted)
 proper frequency. Assuming the end of dilaton-driven inflation to be
quickly followed by the standard radiation-dominated era (which
certainly needs not be the case), one finds $\omega_1\equiv a_1H_1/a
\simeq 10^{11}(H_1/M_p)^{1/2} Hertz$ .
Eq. (\ref{spectra})
 can be compared to the CMB spectrum:
\bq
\frac{d \rho_{cmb}}{d \ln \omega} \simeq
 \omega^4 \left(e^{\frac{\omega}{T_{\g}}}-1\right)^{-1} \; .
\label{CMB}
\eq

Modulo logarithms, the two  spectra agree with each other
 but, of course, there
is no reason to expect $T_{\gamma} \simeq 2.7 K$
 to be very close to $\omega_1$
 since gravitons decoupled very early from everything else while
photons underwent a complicated history until decoupling.
 Yet, amusingly enough, if $H_1\sim M_p$,
the expected value of $T_{gr}\equiv \omega_1$ (under the assumption
of a quick transition to radiation dominance)
 is of the same order as $T_{\gamma}$.

Like any Planckian spectrum, our graviton-dilaton spectrum is also
strongly tilted towards large wavenumbers with a spectral index
$n=4$,
in contrast to the de-Sitter case ($n=1$) and in
agreement with previous computations on the  rate of graviton
and dilaton production in a string cosmology context \cite{GV3,gas1}.
These tilted spectra contain most of their total power
near the maximal proper frequency $\omega_1$.
It would be an interesting challenge to conceive experimental
apparatuses
able to detect a
relic stochastic gravitational background of such an intensity  and
in such a high-frequency range.

\vfill\eject
\renewcommand{\theequation}{A.\arabic{equation}}
\setcounter{equation}{0}
\noindent
{\bf \grande Appendix
\vskip 5mm
 Scalar perturbations
in higher-dimensional
backgrounds}

We shall first discuss the growth of scalar perturbations, in the
longitudinal gauge, for isotropic, dilaton-driven backgrounds with
$d>3$ spatial dimensions. In $d$ dimensions eq.
(\ref{psieq}) is modified: by combining the $d$-dimensional
generalization of the perturbation equations
(\ref{pert1})-(\ref{constraint})
and of the
background equations (\ref{background})
we get
for $\psi_{k}$ (see for instance \cite{GV3})
%%%%%%%%%%%%%%%%%%%%%%%%%%%%%%%%%%%%%%%%%%%%%%%%%%%%%%%%%%
\bq
 \psi''_k +3(d-1) \H \psi'_k +k^2 \psi_k=0
\label{psievolution}
\eq
%%%%%%%%%%%%%%%%%%%%%%%%%%%%%%%%%%%%%%%%%%%%%%%%%%%%%%%%%%
which, for $|k\eta|<<1$, has the asymptotic solution
%%%%%%%%%%%%%%%%%%%%%%%%%%%%%%%%%%%%%%%%%%%%%%%%%%%%%%%%%%
\bq
 \psi_k =c_1+ c_2 \frac{\n}{a^{3(d-1)}}
\eq
%%%%%%%%%%%%%%%%%%%%%%%%%%%%%%%%%%%%%%%%%%%%%%%%%%%%%%%%%%
In $d$-dimensions also the inflationary background solution
is modified \cite{GV3}
%%%%%%%%%%%%%%%%%%%%%%%%%%%%%%%%%%%%%%%%%%%%%%%%%%%%%%%%%%
\bq
 a=(-\n)^{1/(d-1)},\hspace{.5in}
\vphi=-\sqrt{2d(d-1)}\  \ln a
\eq
%%%%%%%%%%%%%%%%%%%%%%%%%%%%%%%%%%%%%%%%%%%%%%%%%%%%%%%%%%
As a consequence, the scalar mode is still growing asymptotically
as
$\eta^{-2}$, exactly as in $d=3$.
 The normalized spectral amplitude acquires  however a
$d$-dependence,
since the expression of $\psi$ in terms of the variable $v$
satisfying canonical commutation relations
becomes \cite{phd}
%%%%%%%%%%%%%%%%%%%%%%%%%%%%%%%%%%%%%%%%%%%%%%%%%%%%%%%%%%
\bq
 \psi_k=\frac{1}{2(d-1) k^2} \frac{\vphi'^2}{\H}
\left(\frac{v_k}{z}\right)',
\hspace{.5in} z=\frac{\vphi'}{\H} a^{(d-1)/2 }
\label{ddimpsi}
\eq
%%%%%%%%%%%%%%%%%%%%%%%%%%%%%%%%%%%%%%%%%%%%%%%%%%%%%%%%%%
where
%%%%%%%%%%%%%%%%%%%%%%%%%%%%%%%%%%%%%%%%%%%%%%%%%%%%%%%%%%
\bq v= a^{(d-1)/2}\left( \chi+\frac{\vphi'}{\H}\right), \hspace{.3in}
 v''_k+ \left(k^2-\frac{z''}{z}\right) v_k=0,
\hspace{.3in}\frac{z''}{z}=\frac{(d-1)^2}{4(d-2)}\ \frac{a''}{a}
\eq
%%%%%%%%%%%%%%%%%%%%%%%%%%%%%%%%%%%%%%%%%%%%%%%%%%%%%%%%%%
The solution for $v_{k}$ in the small $|k\eta|$ limit,
%%%%%%%%%%%%%%%%%%%%%%%%%%%%%%%%%%%%%%%%%%%%%%%%%%%%%%%%%%
\bq
 v_k= c_1 z + c_2 z \int^\n\frac{d\n'}{z^2(\n')}
\eq
%%%%%%%%%%%%%%%%%%%%%%%%%%%%%%%%%%%%%%%%%%%%%%%%%%%%%%%%%%
corresponds then to the normalized asymptotic behaviour
%%%%%%%%%%%%%%%%%%%%%%%%%%%%%%%%%%%%%%%%%%%%%%%%%%%%%%%%%%
\bq
 |v_k|= \frac{z}{z_{\rm HC}} \frac{|\ln(-k\n)|}{\sqrt{k}}
\eq
%%%%%%%%%%%%%%%%%%%%%%%%%%%%%%%%%%%%%%%%%%%%%%%%%%%%%%%%%%
which inserted into eq. (\ref{ddimpsi}) leads to the typical
fluctuation amplitude, on scales $k^{-1}$,
%%%%%%%%%%%%%%%%%%%%%%%%%%%%%%%%%%%%%%%%%%%%%%%%%%%%%%%%%%
\bq
    \left|\d _{\psi_k}(\n)\right| \simeq
\left(\frac{H_1}{M_p}\right)^{(d-1)/4} \frac{(k\n_1)^{d/2}}{ (k\n)^2}
\eq
%%%%%%%%%%%%%%%%%%%%%%%%%%%%%%%%%%%%%%%%%%%%%%%%%%%%%%%%%%
The condition $ | \delta_{\psi}| \laq 1$ implies
%%%%%%%%%%%%%%%%%%%%%%%%%%%%%%%%%%%%%%%%%%%%%%%%%%%%%%%%%%
\bq
 \left|\frac{\n}{\n_1}\right|
{\ \lower-1.2pt\vbox{\hbox{\rlap{$>$}\lower5pt\vbox{\hbox{$\sim$}}}}
\ } \left(\frac{H_1}{M_p}\right)^{(d-1)/4}
\left(\frac{k}{k_1}\right)^{(d-4)/4} \eq
%%%%%%%%%%%%%%%%%%%%%%%%%%%%%%%%%%%%%%%%%%%%%%%%%%%%%%%%%%
which is satisfied, if the inflationary evolution is switched off at
a
scale
$H_{1}\sim (a_{1}\eta_{1})^{-1}\le M_{P}$, for all $d\ge 4$. Though
in
a higher
dimensional background the presence of the growing mode is not
eliminated, the
use of the linearized metric perturbation approach is nevertheless
allowed.

The growing mode, which
has
been shown to be present with the same time-dependence for any number
of spatial dimensions (see eq. (\ref{psievolution})), cannot be
eliminated
even by relaxing the isotropy assumption. In order to discuss this
point we shall consider scalar perturbations in the anisotropic
background (\ref{ani}),  (\ref{ddimback}), by setting
(in the longitudinal
gauge),
%%%%%%%%%%%%%%%%%%%%%%%%%%%%%%%%%%%%%%%%%%%%%%%%%%%%%%%%%%
\bq
\delta\vphi =\chi, \hspace{.3in}
\d g_{00}=2 a^2 \phi,\hspace{.3in}
\d g_{ij}=2 a^2 \psi \d_{ij},\hspace{.3in}
\d g_{mn}=2 b^2\xi \d_{mn}
\eq
%%%%%%%%%%%%%%%%%%%%%%%%%%%%%%%%%%%%%%%%%%%%%%%%%%%%%%%%%%
with $\da g_{0m}=0=\da g_{im}$. This choice is certainly justified
if,
in a dimensional reduction context, we
consider
perturbations which are only function of time and of the external
coordinates
 $x^{i}$, $i=1,...,d$. The ($i,j\ne i$) component of the perturbed
Einstein equations gives then a relation between the three
perturbation
variables
%%%%%%%%%%%%%%%%%%%%%%%%%%%%%%%%%%%%%%%%%%%%%%%%%%%%%%%%%%
\bq
\phi=(d-2)\psi+n\xi
\eq
which allows us to eliminate $\phi$ everywhere in the perturbation
equations.
The ($0,i$) components give the constraint
%%%%%%%%%%%%%%%%%%%%%%%%%%%%%%%%%%%%%%%%%%%%%%%%%%%%%%%%%%
\brr
(d-1)\psi'+(d-2)\psi[(d-1)\H+n\F]&+&\nonumber \\
+n\xi'+n\xi[(d-2)\H+(n+1)\F]&=&\hf \vphi'\chi
\err
%%%%%%%%%%%%%%%%%%%%%%%%%%%%%%%%%%%%%%%%%%%%%%%%%%%%%%%%%%
and the ($0,0$) component gives
%%%%%%%%%%%%%%%%%%%%%%%%%%%%%%%%%%%%%%%%%%%%%%%%%%%%%%%%%%
%%%%%%%%%%%%%%%%%%%%%%%%%%%%%%%%%%%%%%%%%%%%%%%%%%%%%%%%%%
\brr
(d-1)\nabla^2\psi-\psi'[d(d-1)\H+n{\cal F}]&+&\nonumber \\
+n\nabla^2\xi
-n\xi'[d\H+(n-1){\cal F}]&=&\hf \vphi'\chi'
\label{anisotroppert1}
\err
%%%%%%%%%%%%%%%%%%%%%%%%%%%%%%%%%%%%%%%%%%%%%%%%%%%%%%%%%%
The perturbation of the dilaton equation (\ref{dilaton}) gives
\bq
\chi''+[(d-1)\H+n {\cal F}]\chi'-\nabla^2\chi=2\vphi'[(d-1)\psi'+n
\xi']
\label{anisotroppert}
\eq
%%%%%%%%%%%%%%%%%%%%%%%%%%%%%%%%%%%%%%%%%%%%%%%%%%%%%%%%%%
Finally, the ($i,i$) and ($m,m$) components of the perturbed Einstein
equations, combined with eq. (\ref{anisotroppert1}), provide the
following
interesting system of coupled equations for the ``external" and
``internal"
perturbations
$\psi$ and $\xi$ \cite{phd}:
%%%%%%%%%%%%%%%%%%%%%%%%%%%%%%%%%%%%%%%%%%%%%%%%%%%%%%%%%%
\brr
&& (d-1)\bl {\,\lower0.9pt\vbox{\hrule \hbox{\vrule height 0.2 cm
\hskip 0.2 cm
  \vrule height 0.2 cm}\hrule}\,}\psi +\psi'\left[3(d-1) \H+ 3 n\F
\right]\br =
\nonumber \\ &&=
-n\bl{\,\lower0.9pt\vbox{\hrule \hbox{\vrule height 0.2 cm \hskip 0.2
cm \vrule
height 0.2 cm}\hrule}\,}\xi +\xi'\left[3(d-1) \H +3 n \F\right]\br;
\nonumber \\ && d\bl {\,\lower0.9pt\vbox{\hrule \hbox{\vrule height
0.2
cm
\hskip 0.2 cm   \vrule height 0.2 cm}\hrule}\,}\psi
+\psi'\left[3(d-1)
\H+
\frac{\F}{d}\left( 2(d-1)(n-1)+nd\right) \right]\br = \nonumber \\
&&=
-(n-1)\bl{\,\lower0.9pt\vbox{\hrule \hbox{\vrule height 0.2 cm \hskip
0.2 cm
\vrule height 0.2 cm}\hrule}\,}\xi +\xi'\left[ \frac{\H}{n-1}\left(3
d
n - d -n
+1\right)+3 n \F\right]\br
\label{anisotper}
\err
%%%%%%%%%%%%%%%%%%%%%%%%%%%%%%%%%%%%%%%%%%%%%%%%%%%%%%%%%%
where $\Box\equiv \frac{{\partial}^2}{\partial{\eta}^2} -{\nabla}^2$.

This system can be easily diagonalized to find the (time-dependent)
linear
combination of $\psi$ and $\xi$ representing the true ``propagation
eigenstates". For our purpose, however, the asymptotic behaviour of
the
modes
$\psi_{k}$, $\xi_{k}$ can be simply obtained by inserting
into  the previous system the ansatz
%%%%%%%%%%%%%%%%%%%%%%%%%%%%%%%%%%%%%%%%%%%%%%%%%%%%%%%%%%
\bq
\psi_k= A (-\n)^x,\hspace{.3in}\xi_k=B(-\n)^x
\eq
%%%%%%%%%%%%%%%%%%%%%%%%%%%%%%%%%%%%%%%%%%%%%%%%%%%%%%%%%%
One then finds from eqs. (\ref{anisotper}) that in the
$|k\eta|\rightarrow 0$
limit there are non trivial solutions for the coefficients $A$ and
$B$
only if
$x=0$ or $x=-2$, which means that, asymptotically,
%%%%%%%%%%%%%%%%%%%%%%%%%%%%%%%%%%%%%%%%%%%%%%%%%%%%%%%%%%
\bq
\psi_k= A_1+\frac{A_2}{ \n^2},\hspace{.3in}
\xi_k=B_1+\frac{B_2}{ \n^2}
\label{solanis}
\eq
%%%%%%%%%%%%%%%%%%%%%%%%%%%%%%%%%%%%%%%%%%%%%%%%%%%%%%%%%%
We thus find for the scalar perturbation modes the same asymptotic
growth, with
the same power-like behaviour in $\eta$, as in the previous case of
$d=3$
isotropic dimensions.

\end{document}